\documentclass[11pt]{article}
\usepackage{latexsym} %\usepackage{epsf}
\usepackage{epsfig,epsf}
\usepackage{graphicx}
\usepackage{amsfonts}
\usepackage{amsmath}
\usepackage{amsthm}
\usepackage{amssymb}
\renewcommand{\theequation}{\thesection.\arabic{equation}}
\newcounter{subequation}[equation]
\makeatletter

\newcommand{\p}{^{\prime}}
\newcommand{\pp}{^{\prime \prime}}

\expandafter\let\expandafter\reset@font\csname reset@font\endcsname

\def\subeqnarray{\arraycolsep1pt
    \def\@eqnnum\stepcounter##1{\stepcounter{subequation}%
        {\reset@font\rm(\theequation\alph{subequation})}}
\jot5mm     \eqnarray}

\makeatother

\def\be{\begin{equation}}
\def\lb#1{\label{#1}}
\def\ee{\end{equation}}
\def\bea{\begin{eqnarray}}
\def\eea{\end{eqnarray}}
\def\ba{\begin{array}}
\def\ea{\end{array}}

\def\la{\lambda}

\def\one#1{#1^{\raise5pt\hbox{$\scriptstyle\!\!\!\!1$}}\,{}}
\def\two#1{#1^{\raise5pt\hbox{$\scriptstyle\!\!\!\!2$}}\,{}}

\def\II{\hbox{{1}\kern-.25em\hbox{l}}}
\makeatletter
\def\binrel@#1{\begingroup
  \setboxz@h{\thinmuskip0mu
    \medmuskip\m@ne mu\thickmuskip\@ne mu
    \setbox\tw@\hbox{$#1\m@th$}\kern-\wd\tw@
    ${}#1{}\m@th$}%
  \edef\@tempa{\endgroup\let\noexpand\binrel@@
    \ifdim\wdz@<\z@ \mathbin
    \else\ifdim\wdz@>\z@ \mathrel
    \else \relax\fi\fi}%
  \@tempa
}
\let\binrel@@\relax
\def\overset#1#2{\binrel@{#2}%
  \binrel@@{\mathop{\kern\z@#2}\limits^{#1}}}
\def\underset#1#2{\binrel@{#2}%
  \binrel@@{\mathop{\kern\z@#2}\limits_{#1}}}
\makeatother
\newfont{\bbd}{msbm10 scaled\magstep1}

\newcommand*{\sign}{\mathop{\mathrm{sign}}\nolimits}
\begin{document}
%%%%%%%%%%%%%%%%%%%%%%%%%%%%%%%%%%%%%%%%%%%%%%%%%%%%%%%%%%%%%%%%%%%%%%%%%%%%%%%%%%%%%%%%%%%%%%%%%%%%%%%%%%%%%%%%%%%%%%%%
\begin{titlepage}

\vspace*{1cm}

\begin{center}
{\LARGE \bf{ Yangian symmetric  correlators}}

\vspace{1cm}

{\large \sf D. Chicherin$^{ab}$\footnote{\sc e-mail:chicherin@pdmi.ras.ru}
and
R. Kirschner$^c$\footnote{\sc e-mail:Roland.Kirschner@itp.uni-leipzig.de} \\
}

\vspace{0.5cm}

\begin{itemize}
\item[$^a$]
{\it St. Petersburg Department of Steklov Mathematical Institute
of Russian Academy of Sciences,
Fontanka 27, 191023 St. Petersburg, Russia}
\item[$^b$]
{\it Chebyshev Laboratory, St.-Petersburg State University,\\
14th Line, 29b, Saint-Petersburg, 199178 Russia}

\item[$^c$]
{\it Institut f\"ur Theoretische
Physik, Universit\"at Leipzig, \\
PF 100 920, D-04009 Leipzig, Germany}
\end{itemize}
\end{center}
\vspace{0.5cm}
\begin{abstract}
Similarity transformations and eigenvalue relations of monodromy operators
composed of  Jordan-Schwinger type $\mathrm{L}$ matrices are considered
and used to define Yangian symmetric correlators of $n$-dimensional
theories. Explicit expressions are obtained and relations are formulated. In
this way basic notions of the Quantum inverse scattering method
provide a convenient formulation for high symmetry and integrability not
only in lower dimensions.
\end{abstract}

\end{titlepage}

\vspace{4cm}

\newpage

%{\small \tableofcontents}
\renewcommand{\refname}{References.}
\renewcommand{\thefootnote}{\arabic{footnote}}
\setcounter{footnote}{0} \setcounter{equation}{0}

\newpage

%%%%%%%%%%%%%%%%%%%%%%%%%%%%%%%%%%%%%%%%%%%%%%%%%%%%%%%%%%%%%%%%%%%%%%%%%%%%%%%%%%%%%%%%%%%%%%%%%%%%%%%%%%%%%%%%%%%%%%%
\section{Construction scheme and motivations}
\setcounter{equation}{0}
We consider $N$-point correlators in $n$ dimensions, i.e.
functions of $n N$ variables interpreted as $N$ points
$\mathbf{x}_i$ with
coordinates $\left(x_{a,i}\right)_{a=1}^{n} = \left( x_{1,i} , x_{2,i} ,
\cdots ,  x_{n,i} \right)$. Associated with these coordinates we shall work
with $n N$ Heisenberg canonical pairs
$\mathbf{p}_i = \left(p_{a,i}\right)_{a=1}^{n}
= \left( p_{1,i} , p_{2,i} , \cdots , p_{n,i} \right)$ and
$\mathbf{x}_i = \left(x_{a,i}\right)_{a=1}^{n} =
\left( x_{1,i} , x_{2,i} , \cdots , x_{n,i} \right)$,
$ [p_{a,i}, x_{b,j}] = \delta_{a,b} \delta_{i,j} $.
We consider the partition of the index set $N = \{1, 2, \cdots , N \}$
labeling these points in two
non-overlapping subsets $I$ and $J$, $I \cup J = N, \,I \cap J = \varnothing$.
The partition of the set of $N$ sites into the subsets $I, J$ can also be denoted
as signature, i.e.
by a sequence of symbols $''+'',\,''-''$ where the symbol $''+''$ is put for a site in
$I$ and $''-''$ for a site in $J$.

We define the  action of $g\ell_n$ on the points in dependence of their
signature as
$$ \delta x_{c,i} = \left[ \left [ \mathrm{L}^{+}_i\right]_{a,b}, \   x_{c,i}
\right ], \ i \in I, \ \ \   \delta x_{c,j} = \left [\left[\mathrm{L}^{-}_j
\right]_{a,b}, x_{c,j} \right ], \ j \in J $$
\be \lb{genL}
\left[\mathrm{L}^{+}_i\right]_{a,b} =  p_{a,i} x_{b,i} \;,\;\;\;
\left[\mathrm{L}^{-}_j\right]_{a,b} =  - x_{a,j} p_{b,j}\,.
\ee
We denote the
inner product by $( \, \cdot \, )$,
\be \label{inner}
(j i) \equiv (i j) \equiv (\mathbf{x}_i \cdot \mathbf{x}_j ) \equiv x_{a,i} x_{a,j}\,.
\ee
and notice that for $i \in I, j\in J $ it is $g\ell_n$ invariant in the
sense
$$ [\mathrm{L}^{+}_i + \mathrm{L}^{-}_j ,  (\mathbf{x}_i \cdot
\mathbf{x}_j )] = 0\,. $$
Monomials of the form
\be \label{Phi}
 \Phi_{I,J} =  \prod_{i \in I , j\in J} (\mathbf{x}_i \cdot \mathbf{x}_j)^{\lambda_{ij}}
\ee
are $g\ell_n$ invariant  and general $g\ell_n$ invariant
correlators are superpositions of such monomials with varying exponents
which can take complex values. In general the coordinates are complex
valued.

Note that the the inner product results in an invariant if the coordinates
of the involved points $\mathbf{x}_i, \mathbf{x}_j$ transform differently,
one by $\mathrm{L}^{+} $ the other by $\mathrm{L}^{-}$. This is the formal
reason for considering the two actions on coordinates and for introducing
the signature for distinction. In applications to scattering amplitudes this
is related to gluon helicity.

We add the remark that the case $n=2$ is special because there is the
additional invariance relation involving
the symplectic form
\be \label{symplectic}
x_{1,k} x_{2,l} - x_{2,k} x_{1,l} = (\mathbf{x}_k
\mathbf{\cdot \varepsilon}  \mathbf{x}_l),
\ee
$$ [\mathrm{L}^{\pm}_k + \mathrm{L}^{\pm}_l ,  (\mathbf{x}_k
\mathbf{\cdot \varepsilon}\mathbf{x}_l )] = \pm (\mathbf{x}_k
\mathbf{\cdot \varepsilon}\mathbf{x}_l ) \ \mathrm{I}.  $$
 This implies that in this case the invariant correlators depend
in general both on the inner products (\ref{inner}) and on the symplectic
products (\ref{symplectic}).

The trace of the matrix $\mathrm{L}^{\pm}_k $ is $(\mathbf{x}_k \cdot
\mathbf{p}_k)+ n$ or $- (\mathbf{x}_k \cdot \mathbf{p}_k) $.
It commutes
with all the matrix elements and generates the $u(1)$ subalgebra in
$g\ell_n$ acting as infinitesimal dilatations on the coordinates.
Let us restrict the discussion to correlators of definite weights, i.e.
eigenfunctions of all the $N$ dilatation operators
$(\mathbf{x}_k \cdot \mathbf{p}_k)$, $k = 1 , \cdots , N$. The monomial (\ref{Phi}) has dilatation
weights $\sum_{j\in J} \la_{ij}$ for $i \in I$ and  $\sum_{i\in I} \la_{ij}$
for $j\in J$ and a generic $g\ell_n$ invariant correlator with these
weights is given by this monomial multiplied with a function of the
cross ratios
$$ X_{ij,kl} = \frac{(ij) (kl)}{(il)(kj)}\,, \,\,\,\,i, k \in I\,, \,\,\,\, j,l \in J \,.$$
By the restriction to  correlators of definite weights at all points the $n$
dimensional space reduces to the corresponding projective space.

Along with the $N$ point correlator we consider the quantum spin chain
 of $N$ sites with the states at site $k$ forming a representation of
$g\ell_n$ of Jordan-Schwinger type
\cite{J,S,HP,GN,BW,H,BL,DTB,Prag09,Prag12}.
This representation is spanned by monomials
 in the
coordinates of the point $\mathbf{x}_k$ with the eigenvalue of the action of
$(\mathbf{x}_k \cdot \mathbf{p}_k)$ coinciding with the weight of the
correlator at this point.
$(\mathrm{L}^{\pm}_k)_{a,b} $ are the generators
of this representation in dependence on the signature.
The weight characterizes the representation which is irreducible for generic
values.

The $\mathrm{L}$ matrices are defined in terms these generator matrices by
adding a spectral parameter being a complex number,
\be \lb{Lpm}
\mathrm{L}^{+}_k(u_k) = u_k + \mathbf{p}_k \mathbf{x}_k \;,\;\;\;
\mathrm{L}^{-}_k(u_k) = u_k - \mathbf{x}_k \mathbf{p}_k\, ,
\ee
or in  component form
\be \lb{LpmComp}
\left[\mathrm{L}^{+}_k(u_k)\right]_{a,b} = u_k \delta_{a,b} + p_{a,k} x_{b,k} \;,\;\;\;
\left[\mathrm{L}^{-}_k(u_k)\right]_{a,b} = u_k \delta_{a,b} - x_{a,k} p_{b,k}\,.
\ee
In our notations (\ref{Lpm}) we omit a symbol of tensor product $\otimes$ for short.
Notations like $\mathbf{x}_i \mathbf{x}_j$ (for example in the definition
(\ref{Lpm})) are not to be confused with
the inner product (\ref{inner}).

Both operators $\mathrm{L}^{\pm}(u)$ respect the
$\mathcal{R}\mathrm{LL}$-relation with Yang's $\mathcal{R}$-matrix,
\be \label{fund}
\mathcal{R}_{ab,ef}(u-v) \, \mathrm{L}_{ec}(u)\,
\mathrm{L}_{fd}(v) = \mathrm{L}_{bf}(v)\,
\mathrm{L}_{ae}(u)\, \mathcal{R}_{ef,cd}(u-v),
\ee
where $a,b, \cdots = 1,\cdots,n$ and $ \mathcal{R}_{ab,cd}(u) =
u\,\delta_{ac}\,\delta_{bd}+\delta_{ad}\,\delta_{bc}.$
This equation also referred to as the fundamental commutation relation
contains in a compact form all  relations of the underlying Yangian symmetry 
algebra $Y(g\ell_n)$.
The $\mathrm{L}$ matrices  are well known as the basic tool of treating 
the spin chain by
the Quantum Inverse Scattering Method (QISM) \cite{Takhtajan,Kulish,KRS,
Faddeev96}.

In the present paper we are going to study symmetry conditions on
correlators going beyond the mentioned $g\ell_n$ invariance to be formulated
in terms of the $\mathrm{L}$ matrices. For this aim we write down the
monodromy matrix related to the chain as
the ordered product of $\mathrm{L}^{\pm}$-operators (\ref{Lpm}) each
refering to one of the sites,
\be \label{monodrom}
\mathrm{T}^{\alpha_1 \cdots \alpha_N}_{1 \cdots N}(u_1,\cdots,u_N) =
\mathrm{L}^{\alpha_1}_1(u_1) \mathrm{L}^{\alpha_2}_2(u_2) \cdots \mathrm{L}^{\alpha_N}_N (u_N)\,,
\ee
where $\alpha_1 \alpha_2 \cdots \alpha_N$
denotes the signature, i.e. $\alpha_k = \pm$ at $k = 1,\cdots,N$.
The lower indices refer to the chains site with the representation (the
quantum space)  where the operators act nontrivially.
We will omit supplementary indices when it does not lead to misunderstandings.

We intend to study the similarity transformation of the monodromy matrix
(\ref{monodrom}) by invariant correlators like (\ref{Phi}).
We shall write the result of the similarity transformation by (\ref{Phi}) in terms of the
monodromy matrix with changed spectral parameters plus a remainder
\be \label{scheme}
\Phi \circ \mathrm{T}(u_1 ,\cdots,u_N) \equiv \Phi^{-1} \mathrm{T}(u_1 ,\cdots,u_N) \,\Phi =
\mathrm{T}(u_1\p ,\cdots,u_N \p) + \hat r\,.
\ee
Some remainder terms vanish at special values of the exponents $\lambda_{ij}$ (\ref{Phi}).
Here we adopt the notation $\Phi \circ \mathrm{T} \equiv \Phi^{-1} \mathrm{T} \,\Phi$.

Solutions of the following eigenvalue relations
for the monodromy matrix can be obtained from
such similarity relations,
\be \label{eigen}
\mathrm{T}(u_1,\cdots,u_N) \cdot \Phi(u_1,\cdots,u_N) = E(u_1,\cdots,u_N) \, \Phi(u_1,\cdots,u_N)\,.
\ee
Here the matrix $\mathrm{T}$ with operator elements acts on the
correlator function resulting in the r.h.s proportional to the unit matrix.

Indeed,
by acting with both sides of (\ref{scheme}) on the basic state which is represented by a
constant function of the variables
$\mathbf{x}_{1}, \cdots , \mathbf{x}_{N}$ and by requiring
 the vanishing of the remainder up to a constant,
$$ \hat r \cdot 1 = \eta $$
by choice of parameters $\lambda_{ij}$  we arrive at the above
eigenvalue relation with
\be
E(u_1,\cdots,u_N) = \mathrm{T}(u_1\p,\cdots,u_N\p) \cdot 1  + \eta =
\prod_{I,J} (u_i\p+1) u_j\p + \eta
\ee
where we take into account that $\mathrm{L}^{+}(u)\cdot 1 = u+1$
and $\mathrm{L}^{-}(u) \cdot 1= u$ (\ref{Lpm}).
%%&&&&&&&&&&&&&&&&&&&&&&&&&&&&&&&&&&&&&&&&&&&&&&&&&&&&&&&&&&&&&&&&66

The eigenvalue relation (\ref{eigen}) provides the formulation of the extended symmetry
condition on the correlators to be studied here.
A {\sl Yangian symmetric
correlator} is defined as a $g\ell_n$ invariant correlator of
definite dilatation weights being a solution of (\ref{eigen}).

In the present paper we focus on regular correlators not involving
distributions. 
The symmetric correlator can be represented graphically by drawing the chain with its
sites $k$, marked by the corresponding spectral parameter $u_k$, dilatation weight
$2 \ell_k$ and signature $\pm$,
 and with lines (links) connecting sites $i, j$ of different signature,
provided the dependence on the  corresponding $g\ell_n$ invariant
$(\mathbf{x}_i \cdot \mathbf{x}_j )$ is non-trivial.
Otherwise the
corresponding line is omitted. In this way we distinguish symmetric
correlators corresponding to more or less connected or even disconnected
graphs.

\begin{figure}[htbp]
\begin{center}
\includegraphics{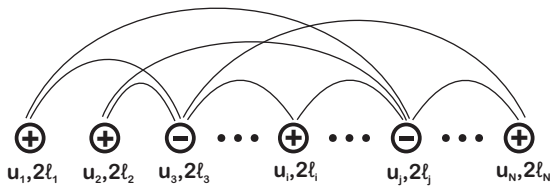}
\caption{Graph representing a symmetric correlator}
\end{center}
\end{figure}

According to the above remark in the special case $n=2$ another type of links between sites of the same
signature  exist. The special features of this case will not be considered
here.

%%%%%%%%%%%%%%%%%%%%%%%%%%%%%%%%%%%%%%%%%%%%%%%%%%%%%%%%%%%
Actually this case is well known because here
the reduction to $s\ell(2)$ can be done in
such a way that the inner products $(\mathbf{x}_i \cdot \mathbf{x}_j )$
and also the symplectic products $(\mathbf{x}_k
\mathbf{\varepsilon} \cdot \mathbf{x}_l) $
turn into
differences in coordinate ratios and that the M\"obius transformations
of those ratios are generated. Symmetric correlators appear e.g. in QCD as
(holomorphic part of) kernels of BFKL  equations \cite{BFKL} of perturbative
Regge asymptotics or of DGLAP/ERBL equations \cite{DGLAP,ERBL} of Bjorken asymptotics.
In the latter case they also generate the anomalous dimensions of composite
operators built of  light-cone components of fields and their derivatives.

The solution of Yang-Baxter $\mathrm{RLL}$ relations by solving the Yangian
conditions on the corresponding kernel (4-point correlator) has been
discussed in \cite{DKK2001}.

The case $n=4$ is related to conformal symmetry in 4 dimensional field
theory. The supersymmetric extension $n= 4|4$ can be formulated
in a straightforward way. Symmetric correlators appear here as kernels of
the renormalization scale dilatation operator or as perturbative scattering
amplitudes in twistor representation.

The observation of the dual conformal and Yangian symmetry of
super Yang-Mills amplitudes attracted great attention
\cite{Drummond,AHCCK1, AHCCK2}. The extended
symmetries  became new
ingredients of the modern tools of amplitude calculations and provide
new insight into the intrinsic structure of gauge theories and their
relation to strings.

The relation of Yang-Mills amplitudes to kernels of Yang-Baxter operators
has been observed in \cite{Ferro} indicating also the potential role of spectral
parameters for regularization of IR divergent loop integrals.
The present approach to Yangian symmetric correlators has been applied to
super Yang-Mills scattering amplitudes in \cite{CDK}.

The notion of  Yangian algebra was introduced by Drinfeld
\cite{Drinfeld} in general form. The case related to $g\ell_n$ was worked
out earlier by L. Faddeev and collaborators \cite{Takhtajan,Kulish,Tarasov} in the QISM
formulation. The relation of the latter formulation to the one in algebra generators
by Drinfeld is  well explained in \cite{Molev}. In the papers discussing the Yangian
symmetry of SYM amplitudes, e.g. \cite{Drummond}  the algebra
generator form is preferred. We use here the advantages of the QISM form.

The plan of the paper is as follows. 
In Section 2 we apply the general scheme outlined above to
construct all $2$- and $3$-point symmetric correlators, the $4$-point correlator for 
the alternating configuration $+-+-$
and the $N+1$-point correlator with the configuration of all signs but one
coinciding.
By a canonical transformation the $2$-point correlators are mapped to 
$\mathrm{R}$-operators.
In Section 3 we proceed to more involved examples
and rewrite the eigenvalue relation for monodromy in the equivalent 
crossing form
that allows to simplify essentially the treatment of the Yangian conditions.
We consider the  signature configurations $-++-$ and $++--$
and express the corresponding correlators in terms of the hypergeometric series 
and in link form.
Further, a generalization of the Yang-Baxter relation is introduced that is intimately
related to the crossing version of the monodromy eigenvalue relation.
In this case the symmetric correlators play the role of kernels of Yang-Baxter operators.
In Section 4 several discrete transformations of the signature configurations
are introduced relating symmetric correlators of equal length.
We consider the reflection of the signature configuration, the mirror transposition, 
the cyclic permutation and the
transposition of a pair $''+''$ and $''-''$.
Then we propose a recurrent procedure that enables one 
to construct higher-point symmetric correlators
sewing lower-point ones with each other as well as to increase the connectivity. 
In Section 5 we summarize.

%%%%%%%%%%%%%%%%%%%%%%%%%%%%%%%%%%%%%%%%%%%%%%%%%%%%%%%%%%%%%%%%%%%%%%%%%%%%%%%%%%%%%%%%%%%%%%%%%%%%%%%%%%%
\section{Monodromy eigenfunctions}
\setcounter{equation}{0}
%%%%%%%%%%%%%%%%%%%%%%%%%%%%%%%%%%%%%%%%%%%%%%%%%%%%%%%%%%%%%%%%%%%%%%%%%%%%%%%%%%%%%%%%%%%%%%%%%%%%%%%%%%%
In this Section we shall show how the general scheme outlined above
works in the cases  of $N=2,3,4$ sites constructing
eigenfunctions of the monodromy matrices (\ref{monodrom})
according to the recipe (\ref{eigen}).

The action of the similarity transformation on a single $\mathrm{L}$-operator 
depends on whether its label falls into the
sets $i \in I$ or $j\in J$,
\be \label{Lsim}
\begin{array}{c}
(\mathbf{x}_i \cdot \mathbf{x}_j)^{-\lambda_{ij}}
\mathrm{L}^{+}_i(u_i) (\mathbf{x}_i \cdot \mathbf{x}_j)^{\lambda_{ij}} =
\mathrm{L}^{+}_i(u_i) + \lambda_{ij} \mathrm{l}_{j\,i}
\\ [0.2 cm]
(\mathbf{x}_i \cdot \mathbf{x}_j)^{-\lambda_{ij}}
\mathrm{L}^{-}_j(u_j) (\mathbf{x}_i \cdot \mathbf{x}_j)^{\lambda_{ij}} =
\mathrm{L}^{-}_j(u_j) - \lambda_{ij} \mathrm{l}_{j\,i}
\end{array}
\ee
where we use the abbreviation
$\mathrm{l}_{ij} \equiv 
\frac{\mathbf{x}_i \mathbf{x}_j}{\left(\mathbf{x}_i \cdot \mathbf{x}_j\right)}$.

Let us note some relations useful in our calculations.
\be \lb{use}
\mathbf{x}_{j} \, \mathrm{l}_{ij} = \mathbf{x}_{j} \;,\;\;\;
\mathrm{l}_{ij} \, \mathbf{x}_{i}= \mathbf{x}_{i} \;,
\ee
$$
\mathrm{l}_{ij}\,\mathrm{l}_{kj} = \mathrm{l}_{ij} \;,\;\;\;
\mathrm{l}_{ij}\,\mathrm{l}_{ik} = \mathrm{l}_{ik} \;,\;\;\;
\mathrm{l}_{ij}\,\mathrm{l}_{ij} = \mathrm{l}_{ij} \;,\;\;\;
\mathrm{l}_{ij}\,\mathrm{l}_{km} =
\frac{(j k)(i m)}{(i j)(k m)} \,\mathrm{l}_{im}
$$

\noindent
{\bf Remark.}
The elementary canonical transformation $\mathtt{C}$ preserving
canonical commutation relations is defined as follows
\be \lb{Canon}
\mathtt{C}^{-1} \,\mathbf{x}\, \mathtt{C} = -\mathbf{p} \;,\;\;\; \mathtt{C}^{-1} \,\mathbf{p}\, \mathtt{C} = \mathbf{x}
\ee
It relates both $\mathrm{L}$-operators to each other
\be \lb{LCanon}
\mathtt{C}^{-1} \,\mathrm{L}^{\pm}(u)\, \mathtt{C} = \mathrm{L}^{\mp}(u)
\ee
The square of canonical transformation changes the signs of coordinate and momentum
$$
\mathtt{C}^{-2} \,\mathbf{x}\, \mathtt{C^2} = -\mathbf{x} \;,\;\;\; \mathtt{C}^{-2} \,\mathbf{p}\, \mathtt{C^2} = -\mathbf{p}
$$

It may be useful to perform the canonical transformation $\mathtt{C}_J$ of the
previous formulae, where $\mathtt{C}_J = \prod_{j\in J} \mathtt{C}_j$ and
$\mathtt{C}_j$ is the canonical transformation in the $j$-th site defined by (\ref{Canon}).
In this case the monodromy matrix (\ref{monodrom}) has to be substituted by the
one constructed out of $\mathrm{L}^{+}$ (see (\ref{LCanon})) only,
\be \lb{monodrom2}
\mathrm{T}^{\alpha_1 \cdots \alpha_N}(u_1,\cdots,u_N) \to
\mathrm{L}^{+}_1(u_1) \cdots \mathrm{L}^{+}_N(u_N)
\ee
The monomial ansatz (\ref{Phi}) transforms to the operator
\be \lb{hatPhi}
\hat\Phi_{I,J} = \prod (\mathbf{x}_i \cdot \mathbf{p}_j)^{\lambda_{ij}}\,.
\ee
The distinction between sites of different signature is shifted now  to the
form of the representation at the sites. The one at a negative signature site $j$
can be described by monomials of $(p_{a,j})$ acting as derivatives on the
distribution $\delta(\mathbf{x}_j)$ representing now the lowest weight state.

This means the action of the canonical transformation on the basic state
can be defined as
\be \lb{basicState}
\mathtt{C} \cdot \delta = 1 \;,\;\;\; \mathtt{C} \cdot 1 = -\delta \;,\;\;\;
\mathtt{C}^{-1} \cdot \delta = - 1 \;,\;\;\; \mathtt{C}^{-1} \cdot 1 = \delta\,.
\ee

%%%%%%%%%%%%%%%%%%%%%%%%%%%%%%%%%%%%%%%%%%%%%%%%%%%%%%%%%%%%%%%%%%%%%%%%%%%%%%%%%%%%%%%%%%%%%%%%%%%%%%%%%%%
\subsection{Two sites}
%%%%%%%%%%%%%%%%%%%%%%%%%%%%%%%%%%%%%%%%%%%%%%%%%%%%%%%%%%%%%%%%%%%%%%%%%%%%%%%%%%%%%%%%%%%%%%%%%%%%%%%%%%%
Let us demonstrate the construction of the eigenfunctions
of the monodromy matrix on the simplest example of two sites.
It is easy to perform the similarity transformation of the monodromy matrix,
$\mathrm{T}^{+-}(u_1,u_2)$ (\ref{monodrom}), using (\ref{Lsim})
$$
(\mathbf{x}_1 \cdot \mathbf{x}_2 )^{\lambda_{12}} \circ \mathrm{L}^{+}_1(u_1)\,
\mathrm{L}^{-}_2(u_2)
%%\, (\mathbf{x}_1 \cdot \mathbf{x}_2 )^{\lambda_{12}} =
%$$
%$$
=\left( \mathrm{L}^{+}_1(u_1+\lambda_{12}) + \lambda_{12}(\mathrm{l}_{21}-1) \right)
\left( \mathrm{L}^{-}_2(u_2-\lambda_{12}) + \lambda_{12}(1-\mathrm{l}_{21}) \right).
$$
Then taking into account the relations 
$\mathbf{x}_1 \mathrm{l}_{21} = \mathbf{x}_1$,
$\mathrm{l}_{21} \mathbf{x}_2= \mathbf{x}_2$, $\mathrm{l}_{21} (\mathrm{l}_{21}-1) = 0$ 
(compare  (\ref{use}))
we  obtain
\be \lb{+-}
(\mathbf{x}_1 \cdot \mathbf{x}_2 )^{\lambda_{12}} \circ
\mathrm{T}^{+-}(u_1,u_2) %%\, (\mathbf{x}_1 \cdot \mathbf{x}_2 )^{\lambda_{12}}
=
\mathrm{T}^{+-}(u_1+\lambda_{12},u_2-\lambda_{12}) +
\lambda_{12} \left( u_1 - u_2 + \lambda_{12} \right) (1-\mathrm{l}_{21})\,.
\ee
At the special value $\lambda_{12}=u_2-u_1$ the remainder in the previous
formula vanishes
\be \lb{+-'}
(\mathbf{x}_1 \cdot \mathbf{x}_2 )^{u_2-u_1} \circ
\mathrm{T}^{+-}(u_1,u_2)
=
\mathrm{T}^{+-}(u_2,u_1)\,.
\ee
Such similarity transformation leads to the permutation of the two spectral
parameters $u_1 \leftrightarrow u_2$.
Then applying both sides of (\ref{+-}) to the vacuum state $1$
one gets the eigenvalue relation for the monodromy matrix
(see (\ref{eigen}))
\be \lb{2point+-}
\mathrm{T}^{+-}(u_1,u_2) \cdot \Phi^{+-} = u_1(u_2+1) \,\Phi^{+-}\;,\;\;\;
\Phi^{+-} = (\mathbf{x}_1 \cdot \mathbf{x}_2)^{u_2-u_1}\,.
\ee

Next we turn to the second possible configuration $-+$ considering
the similarity transformation of the monodromy matrix $\mathrm{T}^{-+}(u_1,u_2)$.
Similar to the previous calculation one  obtains easily
\be \lb{-+}
(\mathbf{x}_1 \cdot \mathbf{x}_2 )^{\lambda_{21}} \circ
\mathrm{T}^{-+}(u_1,u_2)
%%\, (\mathbf{x}_1 \cdot \mathbf{x}_2 )^{\lambda_{21}}
=
\mathrm{T}^{-+}(u_1,u_2) %%+
+ \lambda_{21} \,\mathrm{l}_{12} \left( u_1 - u_2 - \lambda_{21}
- (\mathbf{p}_1 \cdot \mathbf{x}_1 ) - (\mathbf{x}_2 \cdot \mathbf{p}_2 ) \right).
\ee

Contrary to (\ref{+-'}) the spectral parameters stay
untouched and the operator remainder term does not turn to zero for any nonzero $\lambda_{21}$.
Taking $\lambda_{21}=u_1-u_2-n$ and acting with both sides of the latter
equation on the basic state $1$ we find
the eigenvalue relation
\be \lb{2point-+}
\mathrm{T}^{-+}(u_1,u_2) \cdot \Phi^{-+} = u_1(u_2+1) \,\Phi^{-+}\;,\;\;\;
\Phi^{-+} = (\mathbf{x}_1 \cdot \mathbf{x}_2)^{u_1-u_2-n}\,.
\ee
We shall see further that the relations (\ref{+-'}) and (\ref{-+}) are
crucial in our discussion.
The first one implies parameter permutation while in the second
the parameters are untouched.

\noindent
{\bf Remark.}
Performing the canonical transformation (\ref{Canon}) at the second site in (\ref{+-})
%%at $\lambda_{12}=u_2-u_1$
we obtain a Yang-Baxter $\mathrm{RLL}$-relation
\be \lb{R+-}
\mathrm{R}_{12}(u_1-u_2)\,
\mathrm{L}^{+}_1(u_1)\,\mathrm{L}^{+}_2(u_2) =
\mathrm{L}^{+}_1(u_2)\,\mathrm{L}^{+}_2(u_1)\,\mathrm{R}_{12}(u_1-u_2) \;,\;\;\;
\mathrm{R}_{12}(u) = (\mathbf{x}_1 \cdot \mathbf{p}_2)^{u}
\ee
where $\mathrm{R}_{12}(u)$ is $\mathrm{R}$-operator acting in the tensor
product of two infinite-dimensional representations of $g\ell_{n}$.\footnote{
This $\mathrm{RLL}$-relation differs from the fundamental one  (\ref{fund}).
Here the $\mathrm{L}$ operators enter in matrix product, act on different spaces
indicated by subscripts $1,2$ and  the operator $\mathrm{R}_{12}$ acts on
the tensor product space. There both $\mathrm{L}$ act on the same space,
enter in tensor product (expressed by explicit indices) and $\mathcal{R}$
is a $n^2 \times n^2$ matrix.}

After the canonical transformation at the first site
the relation (\ref{-+})
can be also interpreted as a Yang-Baxter $\mathrm{RLL}$-relation.
However in this case
one needs to perform the reduction $g\ell_{n} \to s\ell_{n}$
to cancel the remainder term in (\ref{-+}),
\be \lb{R-+}
\mathrm{R}'_{12}(u_1-u_2)\,
\mathrm{L}^{+}_1(u_1|\ell_1)\,\mathrm{L}^{+}_2(u_2|\ell_2) =
\mathrm{L}^{+}_1(u_1|\ell_2 - \textstyle{\frac{u_1-u_2}{2}})\,
\mathrm{L}^{+}_2(u_2|\ell_1 + \textstyle{\frac{u_1-u_2}{2}})\,\mathrm{R}'_{12}(u_1-u_2) \,,
\ee
$$
\mathrm{R}'_{12}(u) = (\mathbf{p}_1 \cdot \mathbf{x}_2)^{u+2\ell_1 - 2 \ell_2}
$$
In the latter formula $\mathrm{L}^{+}(u|\ell)$ stands for $\mathrm{L}$-operator (\ref{Lpm})
restricted to the space of homogeneous functions of degree $2\ell$.
The operator $\mathrm{R}'_{12}$ (\ref{R-+}) acting on the tensor product of
two irreducible $s\ell_n$ representations
$\mathbb{V}_{\ell_1}\otimes\mathbb{V}_{\ell_2}$ parameterized by
two complex spins $\ell_1$, $\ell_2$ maps it into the space
$\mathbb{V}_{\ell_2 - \frac{u_1-u_2}{2}}\otimes\mathbb{V}_{\ell_1 + \frac{u_1-u_2}{2}}$.
For more details see \cite{Prag12}.

\noindent
{\bf Remark.}
Here we assume that coordinates take complex values,
spectral parameters of the monodromy matrix (\ref{monodrom})
are independent and exponents $\lambda_{ij}$ in the ansatz (\ref{Phi})
are generic complex numbers. However if the coordinate variables are restricted to
real values then taking appropriate limits we can find $\Phi$ (\ref{Phi})
in the form of distribution.
Indeed after appropriate regularization a weak limit \cite{Gelfand} gives
$$
x^{\lambda} \to \delta^{(m)}(x) \;\;\;\; \text{at} \;\;\lambda \to -m-1
$$
where $\delta^{(m)}(x)$ is a $m$-th order derivative of
Dirac $\delta$-function at $m =0 , 1 ,2 ,\cdots$
and $x^{-m-1}\sign(x)$ at $m=-1,-2,\cdots$.
Note that the limit respects the dilatation weight.
Thus the $2$-point eigenfunctions (\ref{2point+-}) and (\ref{2point-+}) turn to
$$
\begin{array}{c}
\Phi^{+-} = \delta^{(m)}(\mathbf{x}_1 \cdot \mathbf{x}_2) \;,\;\;\;
E^{+-} = u_1 (u_2+1)\;,\;\;\; u_2 = u_1-m-1\,,
\\[0.2 cm]
\Phi^{-+} = \delta^{(m)}(\mathbf{x}_1 \cdot \mathbf{x}_2) \;,\;\;\;
E^{-+} = u_1 (u_2+1)\;,\;\;\; u_2 = u_1-n + m +1\,.
\end{array}
$$
The symmetry condition for $\mathrm{L}^{+}_2\mathrm{L}^{+}_2 $ has
non-trivial solutions only with delta distributions. This holds  for all
monodromy operators with all signs coinciding. We do not consider singular
solutions further here. Their role has been discussed in the context of
super Yang-Mills amplitudes \cite{CDK}.

%%%%%%%%%%%%%%%%%%%%%%%%%%%%%%%%%%%%%%%%%%%%%%%%%%%%%%%%%%%%%%%%%%%%%%%%%%%%%%%%%%%%%%%%%%%%%%%%%%%%%%%%%%%
\subsection{Three sites}
%%%%%%%%%%%%%%%%%%%%%%%%%%%%%%%%%%%%%%%%%%%%%%%%%%%%%%%%%%%%%%%%%%%%%%%%%%%%%%%%%%%%%%%%%%%%%%%%%%%%%%%%%%%
In order to demonstrate calculations with  three sites we
quote  here the results for the configurations $+-+$, $++-$, $-+-$.
We will show that corresponding $3$-point eigenfunctions and eigenvalues
in (\ref{scheme}) take the form
\be \lb{3point}
\begin{array}{ll}
E^{+-+} = u_1 (u_2+1) (u_3+1) \;,\;\;\; &
\Phi^{+-+} = (12)^{u_2-u_1} (23)^{u_1-u_3-n}
%%(\mathbf{x}_1 \cdot \mathbf{x}_2)^{u_2-u_1} (\mathbf{x}_2 \cdot \mathbf{x}_3)^{u_1-u_3-n}
\\ [0.2 cm]
E^{++-} = u_1 (u_2+1) (u_3+1) \;,\;\;\; &
\Phi^{++-} = (13)^{u_2-u_1} (23)^{u_3-u_2}
%%(\mathbf{x}_1 \cdot \mathbf{x}_3)^{u_2-u_1} (\mathbf{x}_2 \cdot \mathbf{x}_3)^{u_3-u_2}
\\ [0.2 cm]
E^{-+-} = u_1 u_2 (u_3+1) \;,\;\;\; &
\Phi^{-+-} = (23)^{u_3-u_2} (12)^{u_1-u_3-n}
%%(\mathbf{x}_2 \cdot \mathbf{x}_3)^{u_3-u_2} (\mathbf{x}_1 \cdot \mathbf{x}_2)^{u_1-u_3-n}
\end{array}
\ee
In Subsection \ref{DS} we shall explain that all possible $3$-point configurations
can be deduced from one another by means of discrete symmetry transformations.
%%%%%%%%%%%%%%%%%%%%%%%%%%%%%%%%%%%%%%%%%%%%%%%%%%%%%%%%%%%%%%%%%%%%%%%%%%%%%%%%%%%%%%%%%%%%%%%%%%%%%%%%%%%%%%%%%%%%%
\subsubsection{Configuration $+-+$}
%%%%%%%%%%%%%%%%%%%%%%%%%%%%%%%%%%%%%%%%%%%%%%%%%%%%%%%%%%%%%%%%%%%%%%%%%%%%%%%%%%%%%%%%%%%%%%%%%%%%%%%%%%%%%%%%%%%%%%
We perform the similarity transformation (\ref{scheme}) of the monodromy matrix $\mathrm{T}^{+-+}(u_1,u_2,u_3)$
by $\Phi^{+-+}=(\mathbf{x}_1 \cdot \mathbf{x}_2)^{\lambda_{12}} (\mathbf{x}_2 \cdot \mathbf{x}_3)^{\lambda_{32}}$ in two steps.
On the first step due to (\ref{+-}) we permute $u_1 \leftrightarrow u_2$,
$$
(\mathbf{x}_1 \cdot \mathbf{x}_2)^{\lambda_{12}}\circ\mathrm{T}^{+-+}(u_1,u_2,u_3)
%%\,(\mathbf{x}_1 \cdot \mathbf{x}_2)^{\lambda_{12}}
= \mathrm{T}^{+-+}(u_2,u_1,u_3)\;,\;\;\;
\lambda_{12} = u_2-u_1\,.
$$
Before doing the second similarity transformation we are free to
act with the monodromy matrix on a constant function in first space $1_1$,
$$
\mathrm{T}^{+-+}(u_2,u_1,u_3) \cdot 1_1 = (u_2+1) \,\mathrm{T}^{-+}_{23}(u_1,u_3)
$$
where lower indices $2,3$ of the monodromy matrix on the right-hand side
of the latter relation
refer to quantum spaces of the spin chain where it acts nontrivially.
Thus we have reduced the
problem to a 2-point configuration $-+$ considered above. Applying the second similarity
transformation by means of (\ref{-+}) and acting on $1_{2} 1_{3}$
we obtain finally (\ref{3point}).
%%%%%%%%%%%%%%%%%%%%%%%%%%%%%%%%%%%%%%%%%%%%%%%%%%%%%%%%%%%%%%%%%%%%%%%%%%%%%%%%%%%%%%%%%%%%%%%%%%%%%%%%%%%%%%%%
\subsubsection{Configuration $++-$}
%%%%%%%%%%%%%%%%%%%%%%%%%%%%%%%%%%%%%%%%%%%%%%%%%%%%%%%%%%%%%%%%%%%%%%%%%%%%%%%%%%%%%%%%%%%%%%%%%%%%%%%%%%%%%%%%%
The calculation is analogous to the previous one. The similarity transformation
of the monodromy matrix $\mathrm{T}^{++-}(u_1,u_2,u_3)$
by $\Phi^{++-}=(\mathbf{x}_1 \cdot \mathbf{x}_3)^{\lambda_{13}} (\mathbf{x}_2 \cdot \mathbf{x}_3)^{\lambda_{23}}$
is performed in two steps again.
At first we permute $u_2 \leftrightarrow u_3$ (\ref{+-}),
$$
(\mathbf{x}_2 \cdot \mathbf{x}_3)^{\lambda_{23}}\circ\mathrm{T}^{++-}(u_1,u_2,u_3)
%%\,(\mathbf{x}_2 \cdot \mathbf{x}_3)^{\lambda_{13}}
= \mathrm{T}^{++-}(u_1,u_3,u_2)\;,\;\;\;
\lambda_{23} = u_3-u_2\,.
$$
Then we act on a constant function in second space $1_2$,
$$
\mathrm{T}^{++-}(u_1,u_3,u_2) \cdot 1_2 = (u_3+1) \,\mathrm{T}^{+-}_{13}(u_1,u_2)
$$
reducing the number of spin chain site by one and apply the second similarity
transformation by means of (\ref{+-}) obtaining (\ref{3point}).

The case $-+-$ is trated by analogous steps.

%%%%%%%%%%%%%%%%%%%%%%%%%%%%%%%%%%%%%%%%%%%%%%%%%%%%%%%%%%%%%%%%%%%%%%%%%%%%%%%%%%%%%%%%%%%%%%%%%%%%%%%%%%%%%%%%%
\subsection{Four sites in configuration $+-+-$} \lb{4point}
%%%%%%%%%%%%%%%%%%%%%%%%%%%%%%%%%%%%%%%%%%%%%%%%%%%%%%%%%%%%%%%%%%%%%%%%%%%%%%%%%%%%%%%%%%%%%%%%%%%%%%%%%%%%%%%%%
The pattern to implement the similarity transformation of $\mathrm{T}^{+-+-}$
by
\be \lb{Phi+-+-}
\Phi^{+-+-}=(\mathbf{x}_1 \cdot \mathbf{x}_2)^{\lambda_{12}} (\mathbf{x}_1 \cdot \mathbf{x}_4)^{\lambda_{14}}
(\mathbf{x}_2 \cdot \mathbf{x}_3)^{\lambda_{32}} (\mathbf{x}_3 \cdot \mathbf{x}_4)^{\lambda_{34}}
\ee
is analogous to the above calculations of $3$-point correlation functions.
At first the transpositions
$u_1\leftrightarrow u_2$ and $u_3\leftrightarrow u_4$ are performed due to (\ref{+-})
$$
(\mathbf{x}_1 \cdot \mathbf{x}_2)^{\lambda_{12}} (\mathbf{x}_3 \cdot \mathbf{x}_4)^{\lambda_{34}}\circ
\mathrm{T}^{+-+-}(u_1,u_2,u_3,u_4)
%%\,(\mathbf{x}_1 \cdot \mathbf{x}_2)^{\lambda_{12}} (\mathbf{x}_3 \cdot \mathbf{x}_4)^{\lambda_{34}}
= \mathrm{T}^{+-+-}(u_2,u_1,u_4,u_3)
$$
at $\lambda_{12} = u_2-u_1$, $\lambda_{34} = u_4-u_3$.
Then we apply (\ref{-+})
$$
(\mathbf{x}_2 \cdot \mathbf{x}_3)^{\lambda_{32}}\circ
\mathrm{T}^{+-+-}(u_2,u_1,u_4,u_3)
%%\,(\mathbf{x}_2 \cdot \mathbf{x}_3)^{\lambda_{32}}
= \mathrm{T}^{+-+-}(u_2,u_1,u_4,u_3) + \hat{r}
$$
and notice that at $\lambda_{32}=u_1-u_4-n$ the remainder $\hat{r}$ vanishes
after acting on a basic state in the second and the third spaces $1_2 1_3$
$$
\mathrm{T}^{+-+-}(u_2,u_1,u_4,u_3)\cdot 1_2 1_3 = u_1 (u_4 + 1) \,\mathrm{T}^{+-}_{14}(u_2,u_3)\,.
$$
Finally we permute $u_2\leftrightarrow u_3$ (\ref{+-}),
$$
(\mathbf{x}_1 \cdot \mathbf{x}_4)^{\lambda_{14}}\circ
\mathrm{T}^{+-}_{14}(u_2,u_3)
%%\,(\mathbf{x}_1 \cdot \mathbf{x}_4)^{\lambda_{14}}
= \mathrm{T}^{+-}_{14}(u_3,u_2)
$$
at $\lambda_{14} = u_3-u_2$. Thus we have shown that (\ref{eigen}) takes place with
\be \lb{4point+-+-}
E^{+-+-} = u_1 u_2 (u_3+1) (u_4+1) \;,\;\;\;
\Phi^{+-+-} = (12)^{u_2-u_1} (14)^{u_3-u_2}
(23)^{u_1-u_4-n} (34)^{u_4-u_3}\,.
%%(\mathbf{x}_1 \cdot \mathbf{x}_2)^{u_2-u_1} (\mathbf{x}_1 \cdot \mathbf{x}_4)^{u_3-u_2}
%%(\mathbf{x}_2 \cdot \mathbf{x}_3)^{u_1-u_4-n} (\mathbf{x}_3 \cdot \mathbf{x}_4)^{u_4-u_3}\,.
\ee

We notice that the $4$-point correlator depends on the difference of the
spectral parameters. Therefore the eigenvalue relation (\ref{eigen}) can be decomposed in
five independent relations by introducing a
shift $u_i \to u_i + v$. At equal powers of $v$ we find: the trivial identity at $v^4$,
the $g\ell_{n}$ symmetry condition on $\Phi$ at $v^{3}$, which is fulfilled for arbitrary
exponents $\lambda_{ij}$ in $\Phi$ (\ref{Phi+-+-}),
and  the bilinear in the $g\ell_{n}$ generators
condition at $v^{2}$ is to fix these exponents. From the examples we
expect that no freedom is left. Then the two remaining higher order in generators
conditions arising at $v^{1}$ and $v^{0}$ are fulfilled, which appears as a miracle if regarded without
reference to our result.

Examples of 4-point correlators with other signatures will be considered in
next section after having introduced a convenient transformation of the
eigenvalue condition.

%%%%%%%%%%%%%%%%%%%%%%%%%%%%%%%%%%%%%%%%%%%%%%%%%%%%%%%%%%%%%%%%%%%%%%%%%%%%%%%%%%%%%%%%%%%%%%%%%%%%%%%%%%%
\subsection{One-minus and one-plus configurations}
%%%%%%%%%%%%%%%%%%%%%%%%%%%%%%%%%%%%%%%%%%%%%%%%%%%%%%%%%%%%%%%%%%%%%%%%%%%%%%%%%%%%%%%%%%%%%%%%%%%%%%%%%%%
We consider the monodromy matrix $\mathrm{T}^{+ \cdots + -}_{N\cdots1 0}$ out of
$N+1$ elementary $\mathrm{L}$-operators with the signature $+\cdots+-$.
It is a generalization of the configuration $++-$ considered above.
The similarity transformation (\ref{scheme}) is analyzed and optimized iteratively similar to $++-$.
Firstly we note that due to
\be \lb{N+1}
\prod_{1}^{N} (\mathbf{x}_0 \cdot \mathbf{x}_i)^{\lambda_{i0}}
\circ\mathrm{T}^{+ \cdots + -}_{N\cdots1 0}(u_N,\cdots,u_1,u_0) =
\ee
$$ =
\prod_{2}^{N} (\mathbf{x}_0 \cdot \mathbf{x}_i)^{\lambda_{i0}}
\circ\mathrm{T}^{+ \cdots +}_{N\cdots 2}(u_N,\cdots,u_2)
\underline{(\mathbf{x}_0 \cdot \mathbf{x}_1)^{\lambda_{10}}
\circ\mathrm{L}^{+}_1(u_1) \mathrm{L}^{-}_0(u_0)}
$$
one of the similarity transformation can be easily implemented and
the underlined term is equal to
$\mathrm{L}^{+}_1(u_0) \,\mathrm{L}^{-}_0(u_1)$ at $\lambda_{10}=u_0-u_1$ in view of (\ref{+-}).
Further acting on a vacuum state $1_1$ one
obtains
$$
(u_0+1)\prod_{2}^{N} (\mathbf{x}_0 \cdot \mathbf{x}_i)^{\lambda_{i0}}
\circ\mathrm{T}^{+ \cdots + -}_{N\cdots 2 0}(u_N,\cdots,u_2,u_1)\,
$$
which has the form of (\ref{N+1}) one site less. Continuing the procedure after $N$ steps one obtains
eigenvalue relation
\be \lb{N+1point}
\mathrm{T}^{+ \cdots + -}_{N\cdots1 0}(u_N,\cdots,u_1,u_0) \,\Phi^{+\cdots+-} =
(u_0+1) \cdots (u_1+1) u_N \,\Phi^{+\cdots+-}
\ee
for the $(N+1)$-point correlator
$
\Phi^{+\cdots+-} = \prod_{1}^{N} (\mathbf{x}_0 \cdot \mathbf{x}_i)^{u_0-u_i}\,.
$

In  Subsection \ref{DS} we shall show that the correlator
$\Phi^{-\alpha_1\cdots-\alpha_N}$
with the reflected signature configuration
and the correlators $\Phi^{\alpha'_1\cdots\alpha'_N}$,
where $\alpha'_1\cdots\alpha'_N$ is a cyclic permutations of $\alpha_1\cdots\alpha_N$,
are obtained from $\Phi^{\alpha_1\cdots\alpha_N}$.
Thus knowing (\ref{N+1point}) we can deduce immediately the symmetric
correlators of
arbitrary signature configurations with only one plus or only one minus.

%%%%%%%%%%%%%%%%%%%%%%%%%%%%%%%%%%%%%%%%%%%%%%%%%%%%%%%%%%%%%%%%%%%%%%%%%%%%%%%%%%%%%%%%%%%%%%%%%%%%%%%%%%%
\section{Related symmetry conditions} \lb{RMC}
\setcounter{equation}{0}
%%%%%%%%%%%%%%%%%%%%%%%%%%%%%%%%%%%%%%%%%%%%%%%%%%%%%%%%%%%%%%%%%%%%%%%%%%%%%%%%%%%%%%%%%%%%%%%%%%%%%%%%%%%

%%%%%%%%%%%%%%%%%%%%%%%%%%%%%%%%%%%%%%%%%%%%%%%%%%%%%%%%%%%%%%%%%%%%%%%%%%%%%%%%%%%%%%%%%%%%%%%%%%%%%%%%%%%
\subsection{Yangian algebra generators}
%%%%%%%%%%%%%%%%%%%%%%%%%%%%%%%%%%%%%%%%%%%%%%%%%%%%%%%%%%%%%%%%%%%%%%%%%%%%%%%%%%%%%%%%%%%%%%%%%%%%%%%%%%%
If the solution $\Phi$ of the eigenvalue relation for monodromy matrix
(\ref{eigen}) for the signatures
configuration $\alpha_1 \cdots \alpha_N$ depends on the difference of spectral
parameters only then the related shift symmetry
$u_k \to u_k + v$ leads naturally to a decomposition of the
eigenvalue relation (\ref{eigen}) in powers of the shift parameter $v$.
The pattern reminds the one of perturbative expansion.
Since monodromy matrix and corresponding eigenvalue decompose as follows
$$
\mathrm{T}(u_1+v,\cdots,u_N+v)
= v^N \II + \sum_{k=1}^N v^{N-k} \mathrm{F}^{[k]} \;,\;\;\;
E(u_1+v,\cdots,u_N+v) = v^N + \sum^{N}_{k=1} v^{N-k} E^{[k]}\,,
$$
where $\mathrm{F}^{[k]}$ is of degree $k$ in generators of the $g\ell_n$ algebra,
$$
\mathrm{F}^{[1]} = \sum_k \mathrm{L}^{\alpha_k}_k(u_k) \;;\;\;\;
\mathrm{F}^{[2]} = \sum_{k<m} \mathrm{F}_{k,m}^{[2]}\;,\;\;\;
\mathrm{F}_{k,m}^{[2]} \equiv \mathrm{L}^{\alpha_k}_k(u_k) \mathrm{L}^{\alpha_m}_m(u_m) \;;\;\;\; \cdots\,,
$$
the eigenvalue condition (\ref{eigen})
transforms into the sequence of conditions
\be \lb{part}
\mathrm{F}^{[k]} \, \Phi = E^{[k]} \, \Phi\;,\;\;\; k = 1, \cdots, N\,.
\ee
The partial condition of the first level $k=1$ in (\ref{part})
is fulfilled by the ansatz (\ref{Phi}) since
$$
(\mathrm{L}^{+}_i + \mathrm{L}^{-}_j) (\mathbf{x}_i \cdot \mathbf{x}_j)^{\lambda_{ij}} =
(\mathbf{x}_i \cdot \mathbf{x}_j)^{\lambda_{ij}}\,.
$$
In the generic case the second level $k=2$ condition in (\ref{part}) fixes the exponents and thus the
function $\Phi$. Then the remaining conditions of higher levels $k=3,\cdots,N$ are automatically
fulfilled, although their explicit form is complex with increasing $k$.

Further we shall rewrite the eigenvalue problem (\ref{eigen}) 
in the equivalent form (\ref{eigen'})
on the space of functions of definite homogeneity degree. The corresponding
Yangian conditions of the second level $k=2$ are equivalent however they contain
different number of terms $\mathrm{F}^{[2]}_{k,m}$.
%%Since the eigenvalue problems (\ref{eigen}) and (\ref{eigen'}) are equivalent
%%on the space of functions of definite homogeneity degree corresponding
%%Yangian conditions of the second level $k=2$ are equivalent however they contain
%%different number of terms $\mathrm{F}^{[2]}_{k,m}$.
In particular the conditions
(\ref{eigen'}) are  easier to analyze as it will be demonstrated 
in Subsection \ref{-++-}.

The set
$\mathrm{F}^{[1]}_{a,b},\, \mathrm{F}^{[2]}_{a,b} $
is a particular representation of the Yangian algebra
generators, which will reflect the full algebra in the limit of a
infinitely long chain, $N \to \infty$.

%%%%%%%%%%%%%%%%%%%%%%%%%%%%%%%%%%%%%%%%%%%%%%
\subsection{Related monodromy conditions}
%%%%%%%%%%%%%%%%%%%%%%%%%%%%%%%%%%%%%%%%%%%%%

Here we are going to transform the monodromy eigenvalue relation (\ref{eigen})
in a way that is useful in deriving solutions and is connected
 to the Yang-Baxter $\mathrm{RLL}$-relation
typical for QISM.
We rewrite (\ref{eigen}) by factorizing the monodromy operator in two
factors, the first involving the first $K$ $\mathrm{L}$-matrix factors and
the second involving the remaining ones. Multiplying with the inverse of
this first factor we obtain
\be \lb{eigen2}
\mathrm{T}_{K+1\cdots N}(u_{K+1},\cdots,u_N) \cdot \Phi_{I,J} =
E \, \left(\mathrm{T}_{1\cdots K}(u_1,\cdots,u_K)\right)^{-1} \cdot \Phi_{I,J}\,.
\ee
The inverse of the monodromy matrix can be calculated by the inversion of
the $\mathrm{L}$-matrices using

\be \lb{L-1}
\begin{array}{c}
\mathrm{L}^{+}(u)\,\mathrm{L}^{+}(-u-1-(\mathbf{x}\cdot\mathbf{p})) = u (-u-1-(\mathbf{x}\cdot\mathbf{p}))\,,
\\[0.2 cm]
\mathrm{L}^{-}(u)\,\mathrm{L}^{-}(-u-1+(\mathbf{p}\cdot\mathbf{x})) = u (-u-1+(\mathbf{p}\cdot\mathbf{x}))
\end{array}
\ee
which is checked easily. We take into account that
 $(\mathbf{p}\cdot\mathbf{x})=(\mathbf{x}\cdot\mathbf{p}) + n $
corresponds to the one-dimensional subalgebra
in the decomposition $g\ell_{n}=s\ell_n\oplus u(1)$, and thus  commutes
with $\mathrm{L}^{\pm}$. Therefore the inverses
$\left(\mathrm{L}^{\pm}(u)\right)^{-1}$ are obtained immediately from (\ref{L-1}).

In case of the  expression (\ref{Phi}) for $\Phi_{I,J}$  we have
\be \lb{eigen'}
\mathrm{T}^{\alpha_{K+1}\cdots\alpha_N}_{K+1\cdots N}(u_{K+1},\cdots,u_N) \cdot \Phi_{I,J} =
E\p \, \mathrm{T}^{\alpha_{K}\cdots\alpha_1}_{K\cdots 1}(u_K\p,\cdots,u_1\p) \cdot \Phi_{I,J}\,,
\ee
%\be \lb{eigen'}
%\mathrm{L}^{\alpha_{K+1}}_{K+1}(u_{K+1})\cdots
%\mathrm{L}^{\alpha_{N}}_{N}(u_{N}) \cdot \Phi_{I,J} =
%E\p \, \mathrm{L}^{\alpha_{K}}_{K}(u_{K}\p)\cdots
%\mathrm{L}^{\alpha_{1}}_{1}(u_{1}\p) \cdot \Phi_{I,J}\,,
%\ee
\be \lb{u'}
u_k\p = -u_k-1-\sum_{J} \lambda_{kj} \;\;\; \text{at} \;\; k \in I \;,\;\;\;
u_k\p = -u_k-1 + n + \sum_{I} \lambda_{ik} \;\;\; \text{at} \;\; k \in J \;,\;\;\;
E = E\p \prod_1^{K} u_k u_k\p
\ee
We emphasize
 that the equivalence of  (\ref{eigen2}) and (\ref{eigen'}) holds
not only for the monomial ansatz $\Phi$ (\ref{Phi}) but also if
$\Phi$ is a sum of such monomials with the same degree of homogeneity in
 each coordinate $\mathbf{x}_1,\cdots,\mathbf{x}_N$.

The extension of the monomial ansatz can be written in terms of sums
or in link integral form like in \cite{AHCCK1,AHCCK2}
\be \label{link}
\Phi_{I,J} = \int \phi(c) \exp\left(- \sum c_{ij} \,(\mathbf{x}_i \cdot
\mathbf{x}_j)\right)
\prod \mathrm{d} c_{ij}\,.
\ee
In particular, using the integral formula for the Gamma function
\be \label{Gamma}
y^{\la} = \frac{\Gamma(1+\la)}{2 \pi i} \int_{\mathcal{C}}
\mathrm{d} c \,(-c)^{-\la -1} e^{-c y}, \ee
where the contour $\mathcal{C}$ encircles clockwise the positive real
semi-axis starting
at $+\infty - i \epsilon$, surrounding $0$, and ending at $+\infty + i
\epsilon$,
the monomial ansatz acquires the link form with
$ \phi (c) = \prod c_{ij}^{-1-\la_{ij}} $.

Symmetric correlators being
solutions of the monodromy eigenvalue condition (\ref{eigen'})
can be related to kernels of
integral operators obeying generalized Yang-Baxter relations of the type
\be \label{YBgen}
\mathrm{L}^{\alpha_{K+1}}_{K+1}(v_{K+1})\cdots \mathrm{L}^{\alpha_{N}}_{N}(v_{N}) \,
\hat{\mathrm{R}} =  E_R \,
\hat{\mathrm{R}} \, \mathrm{L}^{\alpha_{K}}_{K}(v_{K})\cdots \mathrm{L}^{\alpha_{1}}_{1}(v_{1})\,.
\ee
Let the operator $\hat{\mathrm{R}}$,
mapping a function of $\mathbf{x}_1, \cdots , \mathbf{x}_{K}$ to a
function of $\mathbf{x}_{K+1}, \cdots , \mathbf{x}_{N}$, 
be represented in integral form with the kernel $\mathrm{R}$,
\be \lb{CorrelInt}
\left[\hat{\mathrm{R}} \cdot \psi\right] (\mathbf{x}_{K+1},\cdots,\mathbf{x}_{N}) =
\int \mathrm{d} \mathbf{x}_1 \cdots \mathrm{d} \mathbf{x}_K \,
\mathrm{R}(\mathbf{x}_1, \cdots , \mathbf{x}_{K} , \mathbf{x}_{K+1} , \cdots , \mathbf{x}_N )\,
\psi(\mathbf{x}_1, \cdots, \mathbf{x}_K)\,.
\ee
We will not specify the integration and impose the only condition that
the integrating by parts by means of
\be \lb{LT}
\int \mathrm{d} \mathbf{x} \,\phi(\mathbf{x}) \mathrm{L}^{\pm}(u)
\psi(\mathbf{x}) =
\int \mathrm{d} \mathbf{x} \,\psi(\mathbf{x}) \mathrm{L}^{\pm T}(u)
\phi(\mathbf{x})
\ee
follows the simple transposition rule
\be \lb{Ltransp}
\left[\mathrm{L}^{\pm T}(u)\right]_{ab} = \
-\left[\mathrm{L}^{\pm}(-u-1)\right]_{ab} \,.
\ee
 We rewrite (\ref{YBgen}) as
\be \lb{YBgen'}
\mathrm{L}^{\alpha_{K+1}}_{K+1}(v_{K+1})\cdots
\mathrm{L}^{\alpha_{N}}_{N}(v_{N}) \cdot \mathrm{R} =
E_R\p \, \mathrm{L}^{\alpha_{K}}_{K}(v_{K}\p)\cdots
\mathrm{L}^{\alpha_{1}}_{1}(v_{1}\p) \cdot \mathrm{R}\,,
\ee
where $E_R\p = (-1)^K E_R ,\, v_k\p = - v_k - 1 , \, k = 1, \cdots ,K$.
The previous relation can be identified with (\ref{eigen'}).

Thus we have shown that the eigenvalue relation for
the monodromy
matrix (\ref{eigen}) can be casted in the form (\ref{eigen'}) which is
the integral kernel condition equivalent
to the operator relation (\ref{YBgen})
being a generalization of the Yang-Baxter equation.

We shall apply (\ref{eigen'}) to derive more 4-point correlators.
Then we consider thoroughly the example of (\ref{YBgen}) at $N=4$,
$K=2$ and $E_{R} = 1$ that corresponds to the Yang-Baxter relation. We will
solve this operator equation for different signature configurations
and show that the solutions are related to the corresponding correlator
calculated.

%%%%%%%%%%%%%%%%%%%%%%%%%%%%%%%%%%%%%%%%%%%%%%%%%%%%%%%%%%%%%%%%%%%%%%%%%%%%%%%%%%%%%%%%%%%%%%%%%%%%%%%%%%%
\subsection{More examples}
\subsubsection{Four site configuration $-++-$} \lb{-++-}
%%%%%%%%%%%%%%%%%%%%%%%%%%%%%%%%%%%%%%%%%%%%%%%%%%%%%%%%%%%%%%%%%%%%%%%%%%%%%%%%%%%%%%%%%%%%%%%%%%%%%%%%%%%
Here we are going to solve the eigenvalue  problem (\ref{eigen}) for the configuration $-++-$.
For simplicity we start with the monomial ansatz
$$
\Phi = (1 2)^{\lambda_{21}} (1 3)^{\lambda_{31}} (2 4)^{\lambda_{24}} (3 4)^{\lambda_{34}}
$$
and rewrite the problem in the form (\ref{eigen'}),
\be \lb{+-to+-}
\mathrm{L}^{+}_3(u_3)\,\mathrm{L}^{-}_4(u_4) \cdot\Phi =
E\p\,\mathrm{L}^{+}_2(u_2\p)\,\mathrm{L}^{-}_1(u_1\p) \cdot\Phi
\ee
where $E = u_1 u_2 u_1\p u_2\p E\p$, $u_1\p = -u_1-1+n+\lambda_{21}+\lambda_{31}$,
$u_2\p = -u_2-1-\lambda_{21}-\lambda_{24}$.
It turns out that conditions on $\Phi$ implied by (\ref{+-to+-}) are
easier to analyze in comparison with initial form of the problem.

The left-hand side of (\ref{+-to+-}) is
$$
\Phi \circ \mathrm{L}^{+}_3(u_3)\,\mathrm{L}^{-}_4(u_4) \cdot 1=
(u_3+1+\lambda_{34} \mathrm{l}_{43} + \lambda_{31} \mathrm{l}_{13})
(u_4-\lambda_{24} \mathrm{l}_{42} - \lambda_{34} \mathrm{l}_{43}) -
\lambda_{34} (1-\mathrm{l}_{43}) =
$$
$$
= (u_3+1)u_4 - \lambda_{34} - \lambda_{24}\lambda_{31} z \, \mathrm{l}_{12}
+ \lambda_{31} (u_4-\lambda_{34}) \, \mathrm{l}_{13}
- \lambda_{24} (u_3+1+\lambda_{34}) \, \mathrm{l}_{42}
+ \lambda_{34} (u_4-u_3-\lambda_{34}) \, \mathrm{l}_{43}\,,
$$
and the right-hand side of (\ref{+-to+-}) is
$$
\Phi \circ \mathrm{L}^{+}_2(u_2\p)\,\mathrm{L}^{-}_1(u_1\p) \cdot 1=
(u_2\p+1+\lambda_{21} \mathrm{l}_{12} + \lambda_{24} \mathrm{l}_{42})
(u_1\p-\lambda_{21} \mathrm{l}_{12} - \lambda_{31} \mathrm{l}_{13}) -
\lambda_{21} (1-\mathrm{l}_{12}) =
$$
$$
= u_1\p (u_2\p+1) - \lambda_{21} - \lambda_{24}\lambda_{31} z \, \mathrm{l}_{43}
- \lambda_{31} (u_2\p+1+\lambda_{21}) \, \mathrm{l}_{13}
+ \lambda_{24} (u_1\p-\lambda_{21}) \, \mathrm{l}_{42}
+ \lambda_{21} (u_1\p-u_2\p-\lambda_{21}) \, \mathrm{l}_{12}\,.
$$
In our calculation we use (\ref{use}):
$\mathrm{l}_{13} \mathrm{l}_{42} = z \,\mathrm{l}_{12}$,
$\mathrm{l}_{42} \mathrm{l}_{13} = z \,\mathrm{l}_{43}$,
$z \equiv \frac{(12)(34)}{(13)(24)}$, $\mathrm{l}_{12} \mathrm{l}_{13} = \mathrm{l}_{13}$, etc.
Since in the previous formulae the cross-ratio $z$ appears and the structures
$1,\,\mathrm{l}_{12},\,\mathrm{l}_{13},\,\mathrm{l}_{42},\,\mathrm{l}_{43}$
are linear independent
the monomial ansatz $\Phi$
can produce only degenerate solutions, i.e. less connected ones where some
exponents  $\lambda_{ij} $ vanish.
The ansatz  has to be generalized to
\be \lb{Phi-++-}
\Phi = (1 2)^{\lambda_{21}} (1 3)^{\lambda_{31}} (2 4)^{\lambda_{24}} (3 4)^{\lambda_{34}}
\sum_{m} b_m  z^m\;,\;\;\; z \equiv \frac{(12)(34)}{(13)(24)}\,,
\ee
which has a definite homogeneity degree in each of the four points as it has been stated above.
Then (\ref{+-to+-}) leads to
$$
(u_3+1)u_4 - \lambda_{34} = E\p \left( u_1\p (u_2\p+1) - \lambda_{21} \right)\,,\,\,\,
\lambda_{34}-u_4 = E\p \left( u_2\p+1+\lambda_{21} \right)\,,\,\,\,
u_3+1+\lambda_{34} = E\p \left( \lambda_{21} - u_1\p \right)\,,
$$
$$
(-\lambda_{31}+m-1)(-\lambda_{24}+m-1) \, b_{m-1} = E\p (\lambda_{21}+m)(u_2\p - u_1\p + \lambda_{21} + m) \, b_{m}\,,
$$
$$
E\p (-\lambda_{31}+m-1)(-\lambda_{24}+m-1) \, b_{m-1} = (\lambda_{34}+m)(u_3 - u_4 + \lambda_{34} + m) \, b_{m}\,.
$$
The two previous relations are consistent if $E\p= \pm 1$;
$\lambda_{21}= \lambda_{34}$ and $u_1\p-u_2\p=u_4-u_3$ or
$\lambda_{21}= u_3-u_4+\lambda_{34}$ and $\lambda_{34}= u_2\p-u_1\p+\lambda_{21}$.
All these possibilities can be analyzed and produce essentially the same solutions.

We take $E\p = 1$, $\lambda_{21}= \lambda_{34}$ and $u_1\p-u_2\p=u_4-u_3$.
Then $\lambda_{24} = u_4-u_2-\lambda_{21}$, $\lambda_{31} = u_1-u_3-n-\lambda_{21}$, $E = u_1 u_2 (u_3+1) (u_4 +1)$ and
coefficients $b_m$ in the correlator $\Phi$ (up to irrelevant multiplier)
$$
b_{m} = \frac{\Gamma(m + u_3-u_1 + n + \lambda_{21} )\Gamma(m+ u_2-u_4 + \lambda_{21})}{\Gamma(m+\lambda_{21}+1)\Gamma(m+u_3-u_4+\lambda_{21}+1)}\,.
$$
Now  we impose the additional restriction $\lambda_{21}=u_4-u_3$
that leads to $b_{m} = 0,\,m<0$ and $\Phi$ (\ref{Phi-++-})
takes the form of a hypergeometric series ${}_2 F_1$.
Convergence of the series (\ref{Phi-++-}) is guaranteed in the region $|z|<1$.
Using the formula (\ref{Gamma})
we obtain the link integral representation
up to irrelevant constant factors,
$$
\Phi =
\int \prod \mathrm{d}c \, e^{-c_{ij} (\mathbf{x}_i \cdot \mathbf{x}_{j})}
(c_{12} c_{34})^{-1-\lambda_{21}} (-c_{13})^{-1 - \lambda_{31}} (-c_{24})^{-1-\lambda_{24}}
\sum_{m\geq 0}  \frac{\Gamma(1+\lambda_{21}+m)}{m!}
\left( \frac{c_{13} c_{24}}{c_{12} c_{34}} \right )^m \,.
$$
The series in the latter formula sums up straightforwardly leading to
\be \lb{Phi-++-link}
\Phi =
\int \prod \mathrm{d}c \,
\frac{e^{-c_{ij} (\mathbf{x}_i \cdot \mathbf{x}_{j})} }
     {(-c_{13})^{1+\lambda_{31}} (-c_{24})^{1+\lambda_{24}} (c_{12} c_{34} - c_{13}c_{24})^{1+\lambda_{21}} }\,,
\ee
where $\lambda_{21} = u_4 - u_3$, $\lambda_{24} = u_3 - u_2$, $\lambda_{31} = u_1 - u_4-n$.

The results of the link integral representations (\ref{Phi-++-link}) are close
to the YM amplitude results \cite{AHCCK1} and are to be compared with the
results of \cite{Ferro}.
%%%%%%%%%%%%%%%%%%%%%%%%%%%%%%%%%%%%%%%%%%%%%%%%%%%%%%%%%%%%%%%%%%%%%%%%%%%%%%%%%%%%%%%%%%%%%%%%%%%%%%%%%%%
\subsubsection{Four site configuration $++--$} \lb{++--}
%%%%%%%%%%%%%%%%%%%%%%%%%%%%%%%%%%%%%%%%%%%%%%%%%%%%%%%%%%%%%%%%%%%%%%%%%%%%%%%%%%%%%%%%%%%%%%%%%%%%%%%%%%%
The eigenvalue problem (\ref{eigen}) for the configuration $-++-$
can be resolved in a similar manner
by rewriting it in the form
\be \lb{--to++}
\mathrm{L}^{-}_3(u_3)\,\mathrm{L}^{-}_4(u_4) \,\Phi =
E\p\,\mathrm{L}^{+}_2(u_2\p)\,\mathrm{L}^{+}_1(u_1\p) \,\Phi\,.
\ee
Here $E = u_1 u_2 u_1\p u_2\p E\p$, $u_1\p = -u_1-1-\lambda_{13}-\lambda_{14}$,
$u_2\p = -u_2-1-\lambda_{23}-\lambda_{24}$.

We find a solution in the form
\be \lb{Phi++--}
\Phi = (1 3)^{\lambda_{13}} (1 4)^{\lambda_{14}} (2 3)^{\lambda_{23}} (2 4)^{\lambda_{24}}
\sum_{m} b_m  z^m\;,\;\;\; z \equiv \frac{(14)(23)}{(13)(24)}\;,\;\;\;
\ee
where $\lambda_{14} = \lambda_{23}$, $\lambda_{13} = u_3-u_1-\lambda_{14}$, $\lambda_{24} = u_4-u_2-\lambda_{14}$,
$b_m = \frac{\Gamma(m+u_2-u_4+\lambda_{14})\Gamma(m+u_1-u_3+\lambda_{14})}
{\Gamma(m+1+\lambda_{14})\Gamma(m+1+u_2-u_3+\lambda_{14})}$
and $E = u_1 u_2 (u_3+1)(u_4+1)$.

Imposing  the additional restriction $\lambda_{14}=u_3-u_2$
 leads to $b_{m} = 0,\,m<0$ and $\Phi$ (\ref{Phi++--})
in the form
\be \lb{Phi++--link}
\Phi =
\int \prod \mathrm{d}c \,
\frac{e^{-c_{ij} (\mathbf{x}_i \cdot \mathbf{x}_{j})} }
     {(-c_{13})^{1+\lambda_{13}} (-c_{24})^{1+\lambda_{24}} (c_{14} c_{23} - c_{13}c_{24})^{1+\lambda_{14}} }\,,
\ee
where $\lambda_{13} = u_2 - u_1$, $\lambda_{14} = u_3 - u_2$, $\lambda_{24} = u_4 - u_3$.

The latter expression can be obtained immediately from the configuration $-++-$ using
the discrete symmetry explained below in Section \ref{DS}.
%%%%%%%%%%%%%%%%%%%%%%%%%%%%%%%%%%%%%%%%%%%%%%%%%%%%%%%%%%%%%%%%%%%%%%%%%%%%%%%%%%%%%%%%%%%%%%%%%%%%%%%%%%%
\subsubsection{Yang-Baxter relation $+-\to-+$}
%%%%%%%%%%%%%%%%%%%%%%%%%%%%%%%%%%%%%%%%%%%%%%%%%%%%%%%%%%%%%%%%%%%%%%%%%%%%%%%%%%%%%%%%%%%%%%%%%%%%%%%%%%%
Let us consider the $\mathrm{RLL}$-relation of the form
\be \lb{YB+-to-+}
\hat{\mathrm{R}}_{12}(u-v)\,\mathrm{L}^{+}_{1}(u)\,\mathrm{L}^{-}_{2}(v) =
\mathrm{L}^{-}_{2}(v)\,\mathrm{L}^{+}_{1}(u)\,\hat{\mathrm{R}}_{12}(u-v)\,.
\ee
where $\mathrm{L}$-operators are defined in (\ref{Lpm}). Lower indices
refer to quantum spaces where operators act nontrivially. We look for
the $\hat{\mathrm{R}}$-operator in the integral form
\be \lb{Rint}
\left[\hat{\mathrm{R}} \cdot \psi\right] (\mathbf{x}_{1},\mathbf{x}_{2}) =
\int \mathrm{d} \mathbf{x}_3 \mathrm{d} \mathbf{x}_4 \,
\mathrm{R}(u-v|\mathbf{x}_1, \mathbf{x}_{2} ; \mathbf{x}_{3} , \mathbf{x}_4 )\,
\psi(\mathbf{x}_3, \mathbf{x}_4)\,.
\ee
After integration by parts in (\ref{YB+-to-+}) using (\ref{LT}), (\ref{Ltransp})
the condition on the kernel arises
\be \lb{YB+-to-+'}
\left[\mathrm{L}^{+}_3(-u-1)\mathrm{L}^{-}_4(-v-1) -
\mathrm{L}^{-}_2(v)\mathrm{L}^{+}_1(u)\right] \mathrm{R}(u-v) = 0\,.
\ee
Since the kernel $\mathrm{R}$ depends on the difference of spectral parameters
we can easily separate $u+v$ and $u-v$ in (\ref{YB+-to-+'}) producing two
independent conditions. Thus the defining condition decomposes into the
$g\ell_n$ symmetry condition from the terms proportional to $u+v$
$$
\left[ \mathrm{L}^{+}_1 + \mathrm{L}^{-}_2 + \mathrm{L}^{+}_3 + \mathrm{L}^{-}_4 - 2 \right] \mathrm{R}(u) = 0
$$
where $\mathrm{L}^{\pm} \equiv \mathrm{L}^{\pm}(u = 0)$ (\ref{genL}),
and the Yangian condition
\be \lb{Ycond}
\left[ \mathrm{Y}^{+-}_{34}(u) - \mathrm{Y}^{-+}_{21}(u) \right] \mathrm{R}(u) = 0
\ee
from the terms proportional to $u-v$. Here we use the short-hand notation
\be \lb{Y}
\mathrm{Y}^{\alpha \beta}_{ij}(u) \equiv
\mathrm{L}^{\alpha}_i \mathrm{L}^{\beta}_j + \frac{u-1}{2} \, \mathrm{L}^{\alpha}_i -
\frac{u+1}{2} \, \mathrm{L}^{\beta}_j\,.
\ee
The former condition is satisfied by our ansatz
\be \lb{Rans}
\mathrm{R}(u) =
(12)^{\lambda} (14)^{\bar\mu}
(23)^{\mu} (34)^{\bar\lambda}
%(\mathbf{x}_1 \cdot \mathbf{x}_2)^{\lambda} (\mathbf{x}_1 \cdot \mathbf{x}_4)^{\bar\mu}
%(\mathbf{x}_2 \cdot \mathbf{x}_3)^{\mu} (\mathbf{x}_3 \cdot \mathbf{x}_4)^{\bar\lambda}
\ee
with four arbitrary parameters $\lambda,\bar\lambda,\mu,\bar\mu$ to be fixed by
the
Yangian condition quadratic in generators $\mathrm{L}^{\pm}$.

Substituting our ansatz (\ref{Rans}) in (\ref{Ycond}) we get
$$
(u-\bar{\lambda}) + \lambda (n + u + \lambda + \mu + \bar{\mu})\,\mathrm{l}_{21} +
\mu (u-\bar\lambda)\,\mathrm{l}_{23} + \bar\mu (u-\bar\lambda)\,\mathrm{l}_{41} +
\bar\lambda (u -\bar\lambda)\,\mathrm{l}_{43} =0\,.
$$
Since the structures $1,\,\mathrm{l}_{21},\,\mathrm{l}_{23},\,\mathrm{l}_{41}\,,\mathrm{l}_{43}$
are linear independent we conclude that the nondegenerate solution,
i.e. where none of the exponents vanish, is
(here we have made the substitution $u\to u-v$)
$$
\bar\lambda = u-v \;,\;\;\; \lambda + \bar\lambda + \mu + \bar\mu = - n\,.
$$
Let us note that in the above calculation the cross-ratio $\frac{(12)(34)}{(14)(23)}$
appearing by the structure $\mathrm{l}_{21}$ due to differentiations is safely canceled that
makes the ansatz (\ref{Rans}) valid.

In order to establish the connection between the obtained solution of the
Yang-Baxter equation (\ref{YB+-to-+}) and the $4$-point correlator (\ref{4point+-+-})
for signature configuration $+-+-$ we invert $\mathrm{L}_1$ and $\mathrm{L}_2$ in
(\ref{YB+-to-+}) by means of (\ref{L-1}),
$$
\mathrm{L}^{+}_1\left(-u-1-(\mathbf{x}_1 \cdot \mathbf{p}_1)\right)
\mathrm{L}^{-}_2\left(-v-1+(\mathbf{p}_2 \cdot \mathbf{x}_2)\right)
\mathrm{L}^{+}_3(-u-1)\,\mathrm{L}^{-}_4(-v-1) \, \mathrm{R}(u-v) =
$$
$$
= u v \left(-u-1-(\mathbf{x}_1 \cdot \mathbf{p}_1)\right)
      \left(-v-1+(\mathbf{p}_2 \cdot \mathbf{x}_2)\right) \mathrm{R}(u-v)\,.
$$
The latter formula is equivalent to (\ref{4point+-+-}) with
$u_1 = \mu+n-v-1,\,u_2 = \lambda + \mu + n-v-1,\, u_3 = -u -1,\, u_4 = -v -1$.
%%$$
%%\mathrm{L}^{+}_1(\mu+n-v-1)
%%\mathrm{L}^{-}_2(\lambda + \mu + n-v-1)
%%\mathrm{L}^{+}_3(-u-1)\mathrm{L}^{-}_4(-v-1) \, \mathrm{R}(u-v) =
%%u v (\mu+n-1) (\lambda + \mu + n-v-1) \mathrm{R}(u-v)\,.
%%$$
%%%%%%%%%%%%%%%%%%%%%%%%%%%%%%%%%%%%%%%%%%%%%%%%%%%%%%%%%%%%%%%%%%%%%%%%%%%%%%%%%%%%%%%%%%%%%%%%%%%%%%%%%%%
\subsubsection{Yang-Baxter relation $--\to++$} \lb{YB-->++}
%%%%%%%%%%%%%%%%%%%%%%%%%%%%%%%%%%%%%%%%%%%%%%%%%%%%%%%%%%%%%%%%%%%%%%%%%%%%%%%%%%%%%%%%%%%%%%%%%%%%%%%%%%%
Next we consider the $\mathrm{RLL}$-relation
\be \lb{YB--to++}
\hat{\mathrm{R}}_{12}(u-v)\,\mathrm{L}^{-}_{1}(u)\,\mathrm{L}^{-}_{2}(v) =
\mathrm{L}^{+}_{2}(v)\,\mathrm{L}^{+}_{1}(u)\,\hat{\mathrm{R}}_{12}(u-v)\,,
\ee
with $\hat{\mathrm{R}}$-operator in the integral form (\ref{Rint}).
Integrating by parts in (\ref{YB--to++}) one obtains condition on the kernel
\be \lb{YB--to++'}
\left[\mathrm{L}^{-}_3(-u-1)\mathrm{L}^{-}_4(-v-1) -
\mathrm{L}^{+}_2(v)\mathrm{L}^{+}_1(u)\right] \mathrm{R}(u-v) = 0\,.
\ee
The latter relation is a particular case of (\ref{--to++})
at $E\p=1$, $u_3=-u-1$, $u_4=-v-1$ and arbitrary $u_1, u_2$.
Consequently the kernel $\mathrm{R}(u)$
has the form (\ref{Phi++--}) (or (\ref{Phi++--link})).

Equation (\ref{YB--to++'}) can also be solved separating the dependence
on $u+v$ and $u-v$
and getting the $g\ell_n$ symmetry condition and the Yangian condition (see (\ref{Y}))
$$
\left[ \mathrm{L}^{+}_1 + \mathrm{L}^{+}_2 + \mathrm{L}^{-}_3 + \mathrm{L}^{-}_4 - 2 \right] \mathrm{R}(u) = 0
\;,\;\;\;
\left[ \mathrm{Y}^{--}_{34}(u) - \mathrm{Y}^{++}_{21}(u) \right] \mathrm{R}(u) = 0\,.
$$

%%%%%%%%%%%%%%%%%%%%%%%%%%%%%%%%%%%%%%%%%%%%%%%%%%%%%%%%%%%%%%%%%%%%%%%%%%%%%%%%%%%%%%%%%%%%%%%%%%%%%%%%%%%
\section{Relations between symmetric correlators}
%%%%%%%%%%%%%%%%%%%%%%%%%%%%%%%%%%%%%%%%%%%%%%%%%%%%%%%%%%%%%%%%%%%%%%%%%%%%%%%%%%%%%%%%%%%%%%%%%%%%%%
\setcounter{equation}{0}
%%%%%%%%%%%%%%%%%%%%%%%%%%%%%%%%%%%%%%%%%%%%%%%%%%%%%%%%%%%%%%%%%%%%%%%%%%%%%%%%%%%%%%%%%%%%%%%%%%%%%%
In this Section we discuss  several ways to generate symmetric
correlators from given ones.
The first methods rely on transformations of the involved monodromy
operators by matrix transposition and inversion based on the corresponding
relations for the $\mathrm{L}$ matrices. The second involves the elementary 
canonical transformation related to Fourier integral.
The third way takes the correlators as
integral operator kernels assuming an integration prescription. Here the
transposition of the $\mathrm{L}$ operators induced by integration by parts matters.

%%%%%%%%%%%%%%%%%%%%%%%%%%%%%%%%%%%%%%%%%%%%%%%%%%%%%%%%%%%%%%%%%%%%%%%%%%%%%%%%%%%%%%%%%%%%%%%%%%%%%%%%%%%
\subsection{Discrete symmetry transformations} \lb{DS}
%%%%%%%%%%%%%%%%%%%%%%%%%%%%%%%%%%%%%%%%%%%%%%%%%%%%%%%%%%%%%%%%%%%%%%%%%%%%%%%%%%%%%%%%%%%%
Given a symmetric correlator by a  solution of the  eigenvalue relation (\ref{eigen})
for the signature configuration
$\alpha_1\cdots\alpha_N$ one can easily obtain symmetric correlators
for other signature configurations
related with the initial one by a discrete symmetry transformation.

Indeed applying matrix transposition of the $\mathrm{L}$-operators 
(see (\ref{LpmComp}))
\be \lb{Lmtransp}
\left[\mathrm{L}^{\pm}(u)\right]_{ba} =
-\left[\mathrm{L}^{\mp}(-u-1)\right]_{ab} \, ,
\ee
in (\ref{eigen}) we have
\be \label{mirror+inv'}
\mathrm{T}^{-\alpha_{N}\cdots -\alpha_{1}}_{N \cdots 1}(-u_N-1,\cdots,-u_1-1)
\cdot \Phi^{\alpha_{1}\cdots\alpha_{N}} =
(-)^N E^{\alpha_{1}\cdots\alpha_{N}} \, \Phi^{\alpha_{1}\cdots\alpha_{N}}\,.
\ee
We obtain that
 the solution of the eigenvalue relation for the monodromy matrix
$\mathrm{T}_{1\cdots N}(u_1,\cdots,u_N)$ with the signature
configuration $-\alpha_{N}\cdots -\alpha_{1}$
(mirror transposition and flipping of the signs) is
\be \label{mirror+inv}
\Phi^{-\alpha_{N}\cdots -\alpha_{1}}(u_1,\cdots,u_N) =
\left.
\Phi^{\alpha_{1}\cdots\alpha_{N}}(-u_N-1,\cdots,-u_1-1)\right|_{\mathbf{x}_{k}\to \mathbf{x}_{N+1-k}}\,,
\ee
\be
E^{-\alpha_{N}\cdots -\alpha_{1}}(u_1,\cdots,u_N) =
(-)^N E^{\alpha_{1}\cdots\alpha_{N}}(-u_N-1,\cdots,-u_1-1)\,.
\ee
As a simple exercise one can  check that
the previous relations do hold for $3$-point
correlators $\Phi^{+-+}$ and $\Phi^{-+-}$ (\ref{3point}).

Due to results obtained in Section \ref{RMC} relying on the inversion
relation of the $\mathrm{L}$ matrices (\ref{L-1})
we find the correlator with the
mirror permutation $\alpha_{N}\cdots\alpha_1$.
Indeed taking in (\ref{eigen'}) $K=N$
we have
\be \lb{mirror'}
 \mathrm{T}^{\alpha_{N}\cdots\alpha_1}_{N\cdots 1}(u_N\p,\cdots,u_1\p)
\cdot \Phi^{\alpha_{1}\cdots\alpha_{N}} =
E^{\alpha_{N}\cdots\alpha_{1}}(u_N\p,\cdots,u_1\p)\,
\Phi^{\alpha_{1}\cdots\alpha_{N}} \,,
\ee
where $u_k\p$ is defined in (\ref{u'}) and
$$
E^{\alpha_{N}\cdots \alpha_{1}}(u_N\p,\cdots,u_1\p) = \left. \prod_1^{N} u_k u_k\p \right/
E^{\alpha_{1}\cdots\alpha_{N}}(u_1,\cdots,u_N)\,.
$$
Combining both previous transformations (\ref{mirror+inv}) and (\ref{mirror'})
we obtain the correlator for the monodromy matrix
$\mathrm{T}_{1\cdots N}(u_1\pp,\cdots,u_N\pp)$ with the signature
configuration $-\alpha_{1}\cdots -\alpha_{N}$ (flipped signs)
\be \lb{inv}
\Phi^{-\alpha_{1}\cdots -\alpha_{N}}(u_1\pp,\cdots,u_N\pp) =
\Phi^{\alpha_{1}\cdots\alpha_{N}}(u_1,\cdots,u_N)\,,
\ee
\be
E^{-\alpha_{1}\cdots -\alpha_{N}}(u_1\pp,\cdots,u_N\pp) =
\left. \prod_1^{N} (u_k+1) u_k\pp \right/
E^{\alpha_{1}\cdots\alpha_{N}}(u_1,\cdots,u_N)\,,
\ee
where
$$
u_k\pp = u_k+n + \sum_{J} \lambda_{kj} \;\;\; \text{at} \;\; k \in I \;,\;\;\;
u_k\pp = u_k - \sum_{I} \lambda_{ik} \;\;\; \text{at} \;\; k \in J\,.
$$

Further we demonstrate the cyclicity property assuming that $\alpha_{1} = +$.
We do this in four steps.
First  we multiply (\ref{eigen}) by the inverse of the  $\mathrm{L}$-operator
in the first space using  (\ref{L-1})
$$
\mathrm{T}^{\alpha_2\cdots\alpha_N}_{2\cdots N}(u_2,\cdots,u_N) \cdot \Phi =
\frac{E}{u_1 u_1\p}\, \mathrm{L}^{+}_1 (u_1\p) \cdot \Phi
$$
where $u_1\p$ is the same as introduced in (\ref{u'}).
Then we perform the matrix transposition (\ref{Lmtransp})
$$
\mathrm{T}^{-\alpha_N\cdots-\alpha_2}_{N\cdots 2}(-u_N-1,\cdots,-u_2-1) \cdot \Phi =
\frac{(-)^N E}{u_1 u_1\p} \, \mathrm{L}^{-}_1(-u_1\p-1) \cdot \Phi\,,
$$
We multiply from the left by the the inverse of $
\mathrm{L}^{-}_1(-u_1\p-1)$ in order to remove this matrix operator from
r.h.s.
$$
\mathrm{T}^{-\,-\alpha_N\cdots-\alpha_2}_{1 N\cdots 2}(-u_1-1+n,-u_N-1,\cdots,-u_2-1) \cdot \Phi =
(-)^N \frac{(u_1+1-n)(u_1\p+1)}{u_1 u_1\p} \,E \cdot \Phi\,,
$$
and apply matrix transposition (\ref{Ltransp}) once more,
\be \lb{cycl}
\mathrm{T}^{\alpha_2 \cdots \alpha_N +}_{2 \cdots N 1}(u_2, \cdots,u_N,u_1-n) \cdot \Phi =
E_{0} \cdot \Phi\;,\;\;\;
E_{0} = \frac{(u_1+1-n)(u_1\p+1)}{u_1 u_1\p} \,E(u_1,\cdots,u_N)\,.
\ee
As a result we observe the cyclicity property of the symmetric correlators:
another symmetric correlator is obtained by cyclic permutation of the points
together
with signatures and spectral parameters, where the flip
from the first site to the last one
is accompanied by a shift in the spectral parameter.
 We see similarities to the crossing relations for scattering
amplitudes, in particular to relations formulated for the
exact S-matrix approach to scattering in 1+1 dimensions
\cite{BKW, ZZ, GhZ}.

%\noindent
%{\bf Remark}.
The relation (\ref{cycl}), where $\alpha_1 = +$, can be rewritten in operator form
\be \lb{cycl'}
\mathrm{T}^{\alpha_2 \cdots \alpha_N +}_{2 \cdots N 1}(u_2, \cdots,u_N,u_1-n) \cdot \Phi =
\frac{(u_1 + 1 - n)\left(u_1+(\mathbf{x}_1\cdot\mathbf{p}_1)\right)}
     { u_1 \left(u_1+1+(\mathbf{x}_1\cdot\mathbf{p}_1)\right)}\,
E(u_1,\cdots,u_N) \cdot \Phi\,.
\ee
At $\alpha_1 = -$ we have instead,
\be \lb{cycl'-}
\mathrm{T}^{\alpha_2 \cdots \alpha_N -}_{2 \cdots N 1}(u_2, \cdots,u_N,u_1-n) \cdot \Phi =
\frac{(u_1 + 1 - n)\left(u_1-(\mathbf{p}_1\cdot\mathbf{x}_1)\right)}
     { u_1 \left(u_1+1-(\mathbf{p}_1\cdot\mathbf{x}_1)\right)}\,
E(u_1,\cdots,u_N) \cdot \Phi\,.
\ee

%%%%%%%%%%%%%%%%%%%%%%%%%%%%%%%%%%%%%%%%%%%%%%%%%%%%%%%%%%%%%%%%%%%%%%%%%%%%%%%%%%%%%%%%%%%%%%%%%%%%%%%%%%%
\subsection{Signature transpositions}
%%%%%%%%%%%%%%%%%%%%%%%%%%%%%%%%%%%%%%%%%%%%%%%%%%%%%%%%%%%%%%%%%%%%%%%%%%%%%%%%%%%%%%%%%%%%%%%%%%%%%%%%%%%

The operation of elementary canonical transformation $\mathtt{C_i}$ (\ref{Canon})
at site $i$
interchanges $\mathrm{L_i}^{\pm}(u) $ with $\mathrm{L_i}^{\mp}(u) $ and
$\Phi (\cdots,\mathbf{x}_i, \cdots) \cdot 1 $ by $ \Phi( \cdots, \mathbf{p}_i, \cdots)
\cdot \delta( \mathbf{x}_i)
$ (\ref{basicState}). In the discussion of this operation we restrict the
related coordinate variables to real values.
Applying this operation at one site $i$ leads from a regular to a singular
correlator. Let us apply the operation to the sites $i,j$ of different
signature. The symmetric correlator with $+$ at $i$ and $-$ at $j$ is
transformed into a symmetric correlator with the transposed signature,
$-$ at $i$ and $+$ at $j$.
The  result is a regular correlator again, a singular one
appears only intermediately.

To perform the transposition
we write the original correlator in link integral form and
$\delta( \mathbf{x}_i) \delta( \mathbf{x}_j) $
in Fourier integrals.

$$ \Phi \to \mathtt{C}_{i,j} \, \Phi = \int \mathrm{d} c \,\mathrm{d} \mathbf{p}_i \mathrm{d} \mathbf{p}_j \, \phi(c) e^{-Q} =
\int \mathrm{d} c \, \phi(c) e^{-Q\p}\,,$$
$$ Q \equiv c_{i_1 j_1} (\mathbf{x}_{i_1} \cdot\mathbf{x}_{j_1})
 - c_{ij} (\mathbf{p}_{i} \cdot\mathbf{p}_{j})
 +i c_{i j_1} (\mathbf{p}_{i} \cdot\mathbf{x}_{j_1})
 +i c_{i_1 j} (\mathbf{x}_{i_1} \cdot\mathbf{p}_{j})
- i (\mathbf{p}_i \cdot \mathbf{x}_i) -i (\mathbf{p}_j \cdot \mathbf{x}_j)\,.
$$
Here $i_1, j_1$ is summed over the ranges $I,J$ respectively
avoiding the fixed values $i, j$. Performing the integrals over
$\mathbf{p}_i, \mathbf{p}_j$ we obtain link integral of the above form with
the quadratic form in the exponential replaced by
$$
Q\p = c_{i\p j\p}\p (\mathbf{x}_{i\p} \cdot \mathbf{x}_{j\p})\,,\,\,\,
c_{i_1 j_1}\p = c_{i_1 j_1} -\frac{ c_{i j_1} c_{i_1 j}}{ c_{ij}}\,,\,\,\,
c_{i_1 i}\p =  \frac{c_{i_1 j}}{c_{ij}} \,,\,\,\,
c_{j j_1}\p = \frac{c_{ij_1}}{c_{ij}} \,,\,\,\,
c_{ji}\p = - \frac{1}{c_{ij}} \,.
$$
The last step of the transposition operation is the change of integration
variables $c_{i\p j\p} \to c_{i\p j\p}\p$, where $i\p$ run over $I$ and $j\p$ run over $J$.
In particular one can check that
the 4-point correlators with the signatures $+-+-$ and $++--$
are related in this way.
In relation to  amplitudes this transposition has been pointed out in
\cite{AHCCK1} presenting also the 4-gluon example.

%%$$$$$$$$$$$$$$$$$$$$$$$$$$$$$$$$$$$$$$$$$$$$$$$$$$$$$$$$$
%%§§§§§§§§§§§§§§§§§§§§§§§§§§§§§§§§§§§§§§§§§§§§§§§§§§§§§§§

%%%%%%%%%%%%%%%%%%%%%%%%%%%%%%%%%%%%%%%%%%%%%%%%%%%%%%%%%%%%%%%%%%%%%%%%%%%%%%%%%%%%%%%%%%%%%%%%%%%%%%%%%%%
\subsection{Recurrent construction by convolution}
%%%%%%%%%%%%%%%%%%%%%%%%%%%%%%%%%%%%%%%%%%%%%%%%%%%%%%%%%%%%%%%%%%%%%%%%%%%%%%%%%%%%%%%%%%%%%%%%%%%%%%%%%%%
Correlation functions can be regarded as kernels of 
integral operators (\ref{CorrelInt}) provided the integration can be defined.
Corresponding examples of the Yang-Baxter operators
have been considered in Section \ref{RMC}.
The subsequent action of these operators is then also defined.
This means that the convolution of kernels by the given integration
prescription results in further correlation functions.
We assume that the integration allows for a simple integration by parts
defining the transposition rule (\ref{LT}), (\ref{Ltransp}).

We impose the condition that the dilatation weight of an integrated point
plus the weight $n$ of the measure adds up to zero.
By this restriction on the dilatation weights the integration becomes compatible
with the symmetry, i.e.
the reduction to the projective space carrying the $s\ell_n$ irreducible
representations
should be  always defined.
With the transposition relation (\ref{Ltransp})
this results in the compatibility of the symmetry conditions
with the convolution; the result of convolution is
a symmetric correlator.

Let us consider two spin chains of the lengths $N$, $\overline{N}$.
The sites of the first one are enumerated by $1,2,\cdots,N$ and of the
second one by $\overline{1},\overline{2},\cdots,\overline{N}$. Let
$\Phi$ and $\overline{\Phi}$ obey corresponding monodromy eigenvalue relations
\be \lb{MbarM}
\mathrm{T}_{1\cdots N}(u_1,\cdots,u_N) \cdot \Phi = E \, \Phi \;, \;\;\;
\mathrm{T}_{\overline{1}\cdots \overline{N}}(u_{\overline{1}},\cdots,u_{\overline{N}}) \cdot \overline{\Phi} =
\overline{E}\, \overline{\Phi}\,.
\ee
Their product is a disconnected symmetric $(N+\overline{N})$-point correlator
related to
  the spin chain of $N+\overline{N}$ sites
\be \lb{N+M}
\mathrm{T}_{1\cdots N , \overline{1} \cdots \overline{N}}(u_1,\cdots,u_N,u_{\overline{1}},\cdots,u_{\overline{N}})
\cdot \Phi \overline{\Phi}
= E \overline{E}  \, \Phi \overline{\Phi}\,.
\ee
We would like to produce a connected correlator by
 constructing the symmetric
 $(N+\overline{N} - K - \overline{K} + \overline{\overline{M}})$-point correlator
which obeys the eigenvalue relation
\be \lb{eigen12}
\mathrm{T}_{\overline{\overline{1}} \cdots \overline{\overline{M}} , K+1 \cdots N ,\overline{K + 1} \cdots \overline{N}}(u_{\overline{\overline{1}}},\cdots,u_{\overline{\overline{M}}},
u_{K+1},\cdots,u_N,u_{\overline{K+1}},\cdots,u_{\overline{N}})
\cdot \Psi = E_{0} \,\Psi\,.
\ee

First we write the involved monodromy operators in two factors, the first
factor involving $K$ (respective $\overline{K}$) factors.
We rewrite both eigenvalue problems (\ref{MbarM}) in the form like in (\ref{eigen2})
by multiplication with the inverse of the first factors on both sides.
Then we multiply the resulting equations and obtain
\be \lb{prprod}
\mathrm{T}_{K+1 \cdots N }(u_{K+1},\cdots,u_N)
\ \mathrm{T}_{\overline{K + 1} \cdots \overline{N}}
(u_{\overline{K+1}},\cdots,u_{\overline{N}})
\cdot \Phi \overline{\Phi} =
\ee
$$
= E \overline{E} \,
\Bigl( \mathrm{T}^{-1}_{1 \cdots K}(u_{1},\cdots,u_K) \cdot\Phi \Bigr)
\Bigl( \mathrm{T}^{-1}_{\overline{1} \cdots \overline{K}}(u_{\overline{1}},
\cdots,u_{\overline{K}}) \cdot\overline{\Phi} \Bigr).
$$
Let $\overline{\overline{\Phi}}$ be a
$(K + \overline{K} + \overline{\overline{M}})$-point correlator. We multiply
both sides of (\ref{prprod}) by
$$
\overline{\overline{\Phi}}(\mathbf{x}_{1},\cdots,\mathbf{x}_{K},
\mathbf{x}_{\overline{1}},
\cdots , \mathbf{x}_{\overline{K}}, \mathbf{x}_{\overline{\overline{1}}}, \cdots,\mathbf{x}_{\overline{\overline{M}}})
$$
where the first $K+ \overline{K}$ points coincide with the corresponding ones of
$\Phi$ and $\overline{\Phi}$
and integrate over these points under the restriction
\be \label{restriction}
[(\mathbf{x}_{l} \cdot \mathbf{p}_{l}) +n ]\  \Phi \, \overline{\Phi}\, \overline{\overline{\Phi}} = 0 \,, \,\,\,
l=1, \cdots,K , \, \overline{1}, \cdots , \overline{K}.
\ee
The resulting correlator
\be \lb{Psi}
\Psi = \int \mathrm{d} \mathbf{x}_1 \cdots \mathrm{d} \mathbf{x}_{K}
\mathrm{d} \mathbf{x}_{\overline{1}} \cdots \mathrm{d} \mathbf{x}_{\overline{K}}
\,\Phi(\mathbf{x}_1,\cdots,\mathbf{x}_K,\cdots)\,
\overline{\Phi}(\mathbf{x}_{\overline{1}},\cdots,\mathbf{x}_{\overline{K}},\cdots)\,
\overline{\overline{\Phi}}(\mathbf{x}_1,\cdots,\mathbf{x}_K,
\mathbf{x}_{\overline{1}},\cdots,\mathbf{x}_{\overline{K}},\cdots)
\ee
obeys
$$
\mathrm{T}_{K+1 \cdots N} (u_{K+1},\cdots,u_N)\,
\mathrm{T}_{\overline{K + 1} \cdots \overline{N}}
(u_{\overline{K+1}},\cdots,u_{\overline{N}})
\cdot \Psi = E\,\overline{E}
\int \mathrm{d} \mathbf{x}_1 \cdots \mathrm{d} \mathbf{x}_{K}
\mathrm{d} \mathbf{x}_{\overline{1}} \cdots \mathrm{d} \mathbf{x}_{\overline{K}}\,
$$
$$
\Bigl( \mathrm{T}^{-1 T}_{1 \cdots K}(u_{1},\cdots,u_K)\,
       \mathrm{T}^{-1 T}_{\overline{1} \cdots \overline{K}}(u_{\overline{1}},\cdots,u_{\overline{K}})
\cdot\overline{\overline{\Phi}}(\mathbf{x}_1,\cdots,\mathbf{x}_K,
\mathbf{x}_{\overline{1}},\cdots,\mathbf{x}_{\overline{K}},\cdots)
\Bigr)
\Phi(\mathbf{x}_1,\cdots,\mathbf{x}_N)
\overline{\Phi}(\mathbf{x}_{\overline{1}},\cdots,\mathbf{x}_{\overline{N}})\,.
$$
This is an intermediate step towards (\ref{eigen12}). Indeed, we act now
with the monodromy operator $\mathrm{T}_{\overline{\overline{1}} \cdots
\overline{\overline{M}}}
(u_{\overline{\overline{1}}},\cdots,u_{\overline{\overline{M}}})$
on the unintegrated points of $\overline{\overline{\Phi}}$ and impose the
condition on the latter correlator to satisfy the eigenvalue relation
\be \lb{monodrAux}
\mathrm{T}_{\overline{\overline{1}} \cdots \overline{\overline{M}}}
(u_{\overline{\overline{1}}},\cdots,u_{\overline{\overline{M}}})\,
\mathrm{T}^{-1 T}_{1 \cdots K}(u_{1},\cdots,u_K)\,
\mathrm{T}^{-1 T}_{\overline{1} \cdots \overline{K}}(u_{\overline{1}},\cdots,u_{\overline{K}})
\cdot \overline{\overline{\Phi}}
= \overline{\overline{E}} \ \overline{\overline{\Phi}}\,.
\ee
Then the intermediate relation turns into (\ref{eigen12}) with the eigenvalue
$
E_0 = E\,\overline{E}\,\overline{\overline{E}}\,.
$
The monodromy operator is composed of three factors
%%$$\mathrm{T}_{\overline{\overline{1}} \cdots \overline{\overline{M}}  ,
%%K+1 \cdots N ,\overline{K} + 1 \cdots \overline{N}}(u_{\overline{\overline{1}}},
%%\cdots,u_{\overline{\overline{M}}},
%%u_{K+1},\cdots,u_N,u_{\overline{K}+1},\cdots,u_{\overline{N}})
%%= $$
$$
\mathrm{T}_{\overline{\overline{1}} \cdots \overline{\overline{M}}}
(u_{\overline{\overline{1}}},\cdots,u_{\overline{\overline{M}}})\,
\mathrm{T}_{K+1 \cdots N }(v_{K+1},\cdots,v_N)\,
\mathrm{T}_{\overline{K + 1} \cdots \overline{N}}
(v_{\overline{K+1}},\cdots,v_{\overline{N}})
$$
where the relation between the spectral parameters $v_m$ and $u_m$ 
can be easily established by means of (\ref{L-1}) and (\ref{Ltransp}).

Let us consider an example. We glue $N$- and $\overline{N}$-point correlators
by means of the $4$-point correlator with signature configuration $+-+-$,
i.e. in the previous formulae $K = 1$, $\overline{K}= \overline{1}$,
$\overline{\overline{M}} = \overline{\overline{2}}$.
We also assume that $''+''$ is assigned to sites $1$, $\overline{\overline{1}}$
and $''-''$ is assigned to sites $\overline{1}$, $\overline{\overline{2}}$.
Applying the crossing relation (\ref{cycl}) several times we can shift the selected points
to the corresponding first position in (\ref{MbarM}),
$\mathbf{x}_{1},\, \mathbf{x}_{\overline{1}}$.
Let us assume that
the crossing is already done and we start from the above form.
Integration by parts leads to
$$
\bigl[\mathrm{L}^{+}_1(u_1)\bigr]^{-1 T}
\bigl[\mathrm{L}^{-}_{\overline{1}}(u_{\overline{1}})\bigr]^{-1 T} =
\frac{\mathrm{L}^{+}_{1}\left(u_1-(\mathbf{p}_1\cdot\mathbf{x}_1)\right)
\mathrm{L}^{-}_{\overline{1}}\left(u_{\overline{1}}+(\mathbf{x}_{\overline{1}}\cdot\mathbf{p}_{\overline{1}})\right)}
 {u_1 \left(-u_1-1+(\mathbf{p}_1\cdot\mathbf{x}_1)\right)
  u_{\overline{1}} \left(-u_{\overline{1}}-1-(\mathbf{x}_{\overline{1}}\cdot\mathbf{p}_{\overline{1}})\right)}
$$
and the auxiliary eigenvalue problem (\ref{monodrAux}) takes the form
$$
\mathrm{L}^{+}_{\overline{\overline{1}}}(u_1)
\mathrm{L}^{-}_{\overline{\overline{2}}}(u_{\overline{1}})
\mathrm{L}^{+}_{1}(u)
\mathrm{L}^{-}_{\overline{1}}(v)\cdot
\overline{\overline{\Phi}} = u_1 u_{\overline{1}} (u+1)(v+1)\, \overline{\overline{\Phi}}
$$
where the symmetric  $4$-point correlator is according to (\ref{4point+-+-})
$$
\overline{\overline{\Phi}} =
(\overline{\overline{1}} \, \overline{\overline{2}})^{u_{\overline{1}}\, - u_1}
(\overline{\overline{1}} \, \overline{1})^{u-u_{\overline{1}}}
(\overline{\overline{2}} \, 1)^{u_1 - v - n}
(1 \, \overline{1})^{v - u}\,.
$$
The spectral parameters $u$ and $v$ should be fixed to
assure (\ref{restriction}) for the  degrees of homogeneity of
$\Phi\,\overline{\Phi}\,\overline{\overline{\Phi}}$.
Thus we obtain a symmetric correlator with eigenvalue $E_0 = E \overline{E}$ by fusing $\Phi$ and $\overline{\Phi} $
with the help of the 4-point correlator,
$$
\Psi = \int \mathrm{d} \mathbf{x}_1 \mathrm{d} \mathbf{x}_{\overline{1}}
\,(\overline{\overline{1}} \, \overline{\overline{2}})^{u_{\overline{1}}\, - u_1}
(\overline{\overline{1}} \, \overline{1})^{u-u_{\overline{1}}}
(\overline{\overline{2}} \, 1)^{u_1 - v - n}
(1 \, \overline{1})^{v - u}\,
\Phi(\mathbf{x}_1,\cdots)\,\overline{\Phi}(\mathbf{x}_{\overline{1}},\cdots)
\,. $$
% \begin{array}{c} \includegraphics{conv} \end{array}
\begin{figure}[htbp]
\begin{center}
\includegraphics{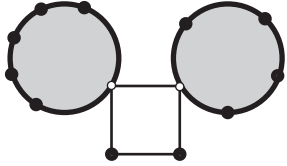}
\caption{Example of correlator convolution }
\end{center}
\end{figure}

%\begin{figure}[htbp]
%\begin{center}
%\vspace{-0.cm}
%\epsfig{conv,width=7.9cm}
%\end{center}
%\caption{Example of correlator convolution }
%\end{figure}

This procedure of connecting solutions has  reminiscence to the BCFW
\cite{BCFW}
prescription for super YM amplitudes in the formulation by integral
\cite{AHCCK1}.

The procedure can be reformulated starting instead of the product
of two correlators from a single not necessarily disconnected
$\Phi$ involving all the considered $N + \overline{N}$ points. The convolution
with $\overline{\overline{\Phi}}$ may increase the connectivity and
 is reminiscent to the loop integration of amplitudes.

\section{Discussion}

Yangian symmetric correlators have been introduced for the aim of
generalizing the known cases of kernels Yang-Baxter operator kernels and for
allowing a general view on the features of Yangian symmetry found in the case
of super Yang-Mills scattering amplitudes.
Indeed, we have encountered here
a number of relations known from this case.

Yangian symmetric correlators are the solutions of the eigenvalue relations
 monodromy operators restricted to definite dilatation weights at the
points. To each of the $n$-dimensional points a dilatation weight, a
spectral parameter and a signature is associated. These features have their
natural origin in the symmetry context, without any reference to the example
of amplitudes.
The individual spectral parameters are useful ingredients; their role is not
restricted to the one of expansion parameter in the monodromy operators
as being generating functions of the Yangian algebra generators.

Comparing to the properties of amplitudes it is clear that fixing the
dilatation weight corresponds to imposing the helicity constraint and that
the signature is related to the gluon helicity. The potential advantage
in  amplitude calculations of
allowing for a spectral parameter dependence has been pointed out  in
\cite{Ferro}.

The Jordan-Schwinger type representations of $g\ell_n$ is a simple basic
case of higher rank ($A_{n-1}$) symmetry. Generic irreducible
representations of this type are characterized by just one parameter
(related to the dilatation weight). Such representation are the building
units by which general representations can be constructed iteratively.
The simplicity is manifest in the simple form of the $\mathrm{L}$-matrices,
leading by elementary steps to relations for similarity transformations,
inversion, matrix transposition and operator conjugation. These transformations
result in relations for monodromy operators and consequently for symmetric
correlators.

The $n$ degrees of freedom associated with any chain site
can be related to a $n$ dimensional position space. In this way of the
applications of the corresponding integrable dynamical system are not 
necessarily restricted to one or
two (discrete) dimensions. The one-dimensional structure of the associated
spin chain is reflected in the cyclicity property of the correlators.

%%%%%%%%%%%%%%%%%%%%%%%%%%%%%%%%%%%%%%%%%%%%%%%%%%%%%%%%%%%%%%%%%%%%%%%%%%%%%%%%%%%%%%%%%%%%%%%%%%%%%%%%%%%%%%%%%%
%\vspace{0.5 cm}

%\hrule

%\vspace{0.5 cm}

\section*{Acknowledgement}
The authors are grateful to Sergey Derkachov for useful discussions. 

The work of D.C. is supported by the Chebyshev Laboratory
(Department of Mathematics and Mechanics, St.-Petersburg State University)
under RF government grant 11.G34.31.0026, by JSC "Gazprom Neft"
and by Dmitry Zimin's "Dynasty" Foundation. He thanks Leipzig University
for hospitality and DAAD for support.


\begin{thebibliography}{99}
%\footnotesize\itemsep=0pt

%% JS %%%
\bibitem{J}
P. Jordan,
``Der Zusammenhang der symmetrischen und linearen Gruppen und
das Mehrk\"orperproblem,''
{\it Zeitschr. f. Physik}, {\bf 94} (1935),
331--335.

\bibitem{S}
J. Schwinger, notes (1952) reprinted in:
Quantum Mechanics of Angular Momentum,
 Biedenharn L.C. and van Dam H. (eds.),
Academic Press, London 1965.



\bibitem{HP}
T. Holstein and H. Primakoff,
Field dependence of the intrinsic domain magnetization of a ferromagnet.
Phys. Rev. {\bf 58} (1940), 1098 - 1113.


\bibitem{GN}I.M. Gelfand  and  M.A. Naimark,
Unitary representations of the classical groups,
Trudy Math. Inst. Steklov, Vol. 36, Moscow-Leningrad 1950.
(German translation: Akademie Verlag, Berlin 1957)

\bibitem{BW}
A. Borel  and A. Weil, Representations lineaires et
espaces homogenes
K\"ahlerians des groupes de Lie compactes,
Sem. Bourbaki, May 1954, (expose J.-P. Serre).


\bibitem{H}
K.T. Hecht, The vector coherent state method and its application
to problems of higher symmetry.
Lecture Notes in Physics 290,
Springer, 1987.


\bibitem{BL}
L.C. Biedenharn  and  M.A. Lohe ,
An extension of the Borel-Weil construction to the
quantum group $U_q(n)$, {\it CMP} {\bf 146} (1992) 483--504;
Quantum group symmetry and q-tensor algebras,
World Scientific 1995.


%\cite{Dobrev:1994ne}
\bibitem{DTB}
V.K. Dobrev, P. Truini and L.C. Biedenharn,
``Representation theory approach to the polynomial solutions of q
difference equations:U-q(sl(3)) and beyond, ''
  J.\ Math.\ Phys.\  {\bf 35} (1994) 6058
  [arXiv:q-alg/9502001];
  %%CITATION = JMAPA,35,6058;%%



%%%%%%%%%%%%%%%%%%%%%%%%%%%%%%%%%%%%%%%%%%%%%%%%%%%%%%%%%%

%%%%%%%%%%%%%%Yangian symmetry in amplitudes   %%%%%%%%%%%%
\bibitem{Drummond}
  J.~M.~Drummond, J.~Henn, G.~P.~Korchemsky and E.~Sokatchev,
  ``Dual superconformal symmetry of scattering amplitudes in N=4
  super-Yang-Mills theory,''
  Nucl.\ Phys.\ B {\bf 828} (2010) 317
  [arXiv:0807.1095 [hep-th]].
  %%CITATION = ARXIV:0807.1095;%%
%\cite{Drummond:2009fd}
%\bibitem{Drummond:2009fd}

 J.~M.~Drummond, J.~M.~Henn and J.~Plefka,
  ``Yangian symmetry of scattering amplitudes in N=4 super Yang-Mills
  theory,''
  JHEP {\bf 0905} (2009) 046
  [arXiv:0902.2987 [hep-th]].
  %%CITATION = ARXIV:0902.2987;%%
%\cite{Drummond:2010qh}
%\bibitem{Drummond:2010qh}

  J.~M.~Drummond and L.~Ferro,
  ``Yangians, Grassmannians and T-duality,''
  JHEP {\bf 1007} (2010) 027
  [arXiv:1001.3348 [hep-th]].
  %%CITATION = ARXIV:1001.3348;%%



%\cite{ArkaniHamed:2009dn}
\bibitem{AHCCK2}
  N.~Arkani-Hamed, F.~Cachazo, C.~Cheung and J.~Kaplan,
  ``A Duality For The S Matrix,''
  JHEP {\bf 1003} (2010) 020
  [arXiv:0907.5418 [hep-th]].
  %%CITATION = ARXIV:0907.5418;%%
  %147 citations counted in INSPIRE as of 11 Mar 2013

%\cite{ArkaniHamed:2009si}
\bibitem{AHCCK1}
  N.~Arkani-Hamed, F.~Cachazo, C.~Cheung and J.~Kaplan,
  ``The S-Matrix in Twistor Space,''
  JHEP {\bf 1003} (2010) 110
  [arXiv:0903.2110 [hep-th]].
  %%CITATION = ARXIV:0903.2110;%%
  %81 citations counted in INSPIRE as of 11 Mar 2013




%\cite{Ferro:2012xw}
\bibitem{Ferro}
  L.~Ferro, T.~Lukowski, C.~Meneghelli, J.~Plefka and M.~Staudacher,
  ``Harmonic R-matrices for Scattering Amplitudes and Spectral
   Regularization,''
Phys.\ Rev.\ Lett.\  {\bf 110} (2013) 121602,
  [arXiv:1212.0850 [hep-th]].
  %%CITATION = ARXIV:1212.0850;%%

%\cite{Chicherin:2013ora}
\bibitem{CDK}
  D.~Chicherin, S.~Derkachov and R.~Kirschner,
  ``Yang-Baxter operators and scattering amplitudes in $\mathcal{N} = 4$
  super-Yang-Mills theory,''
  arXiv:1309.5748 [hep-th].
  %%CITATION = ARXIV:1309.5748;%%


%%%%%%%%%%%%%%%%%%%%%%%%%%%%%%%%%%%%%%%%%%%%%%%%%%%%%%%%%
%%%%%%%%%%%%%%%%%%%%%%%%%%%%%%%%%%%%%%%%%%%%%%

%\cite{Kirschner:2013ila}
\bibitem{Prag12}
  R.~Kirschner,
  ``Integrable chains with Jordan-Schwinger representations,''
  J.\ Phys.\ Conf.\ Ser.\  {\bf 411} (2013) 012018.
  %%CITATION = 00462,411,012018;%%
%\cite{Karakhanyan:2009hx}

\bibitem{Prag09}
  D.~Karakhanyan and R.~Kirschner,
  ``Jordan-Schwinger representations and factorised Yang-Baxter
   operators,''
  SIGMA {\bf 6} (2010) 029
  [arXiv:0910.5144 [hep-th]].
  %%CITATION = ARXIV:0910.5144;%%



%%^^^^^^^^^^^^^^^^^^^^crossing^^^^^^^^^^^^^^^^^^^^^^^^^^^^^^^^^^^^^^^^^^^^

%\cite{Berg:1978sw}
\bibitem{BKW}
  B.~Berg, M.~Karowski and P.~Weisz,
  ``Construction of Green Functions from an Exact S Matrix,''
  Phys.\ Rev.\ D {\bf 19} (1979) 2477.
  %%CITATION = PHRVA,D19,2477;%%
  %143 citations counted in INSPIRE as of 18 Mar 2013

%\cite{Zamolodchikov:1978xm}
\bibitem{ZZ}
  A.~B.~Zamolodchikov and A.~B.~Zamolodchikov,
  ``Factorized s Matrices in Two-Dimensions as the Exact Solutions of
  Certain Relativistic Quantum Field Models,''
  Annals Phys.\  {\bf 120} (1979) 253.
  %%CITATION = APNYA,120,253;%%
  %1072 citations counted in INSPIRE as of 18 Mar 2013


%\cite{Ghoshal:1993tm}
\bibitem{GhZ}
  S.~Ghoshal and A.~B.~Zamolodchikov,
  ``Boundary S matrix and boundary state in two-dimensional integrable
  quantum field theory,''
  Int.\ J.\ Mod.\ Phys.\ A {\bf 9} (1994) 3841
   [Erratum-ibid.\ A {\bf 9} (1994) 4353]
  [hep-th/9306002].
  %%CITATION = HEP-TH/9306002;%%
  %439 citations counted in INSPIRE as of 18 Mar 2013


\bibitem{DGLAP}
V.G. Gribov and L.N. Lipatov, Sov. J. Nucl. Phys.
15(1972)438 \\
L.N. Lipatov,  Yad. Fiz. 20(1974)532     \\
G.~Altarelli and G.~Parisi, Nucl. Phys.
B126(1977)298 \\
Yu.L. Dokshitzer, ZhETF 71(1977)1216

\bibitem{ERBL} V.L. Chernyak and A.R. Zhitnitsky,
JETP Lett 25 (1977) 510;
\\
 A.V. Efremov, A.V. Radyushkin, Theor. Math. Phys. 42 (1980)
  97; Phys. Lett. B94 (1980) 245.
\newline
S.J. Brodsky, G.P. Lepage, Phys. Lett B87 (1979) 359; Phys. Rev. D22 (1980)
2157.


\bibitem{BFKL} L.N. Lipatov, Sov.J.Nucl.Phys. 23(1976)338          \\
               V.S. Fadin, E.A. Kuraev and L.N. Lipatov,
Phys. Lett. 60B(1975)50;
Sov.Phys. JETP 44(1976)443; {\it ibid} 45(1977)199 \\
                Y.Y. Balitski and L.N. Lipatov, Sov.J.Nucl.Phys. 28(1978)882





%\cite{Derkachov:2001sx}
\bibitem{DKK2001}
  S.~E.~Derkachov, D.~Karakhanyan and R.~Kirschner,
  ``Universal R-matrix as integral operator,''
  Nucl.\ Phys.\ B {\bf 618} (2001) 589
  [nlin/0102024 [nlin-si]].
  %%CITATION = NLIN/0102024;%%
  %12 citations counted in INSPIRE as of 21 May 2013




%\cite{Drinfeld:1985rx}
\bibitem{Drinfeld}
  V.~G.~Drinfeld,
 ``Hopf algebras and the quantum Yang-Baxter equation,''
  Sov.\ Math.\ Dokl.\  {\bf 32} (1985) 254
   [Dokl.\ Akad.\ Nauk Ser.\ Fiz.\  {\bf 283} (1985) 1060];
  %%CITATION = SVMDA,32,254;%%
%\bibitem{Drinfeld:1987sy}
 % V.~G.~Drinfeld,
  ``A New realization of Yangians and quantized affine algebras,''
  Sov.\ Math.\ Dokl.\  {\bf 36} (1988) 212.
  %%CITATION = SVMDA,36,212;%%
  %213 citations counted in INSPIRE as of 21 May 2013

%\cite{Takhtajan:1979iv}
\bibitem{Takhtajan}
  L.~A.~Takhtajan and L.~D.~Faddeev,
  ``The Quantum method of the inverse problem and the Heisenberg XYZ
  %model,''
  Russ.\ Math.\ Surveys {\bf 34} (1979) 11
   [Usp.\ Mat.\ Nauk {\bf 34} (1979) 13].
  %%CITATION = RMSUA,34,11;%%
  %324 citations counted in INSPIRE as of 21 May 2013

%\cite{Kulish:1981bi}
\bibitem{Kulish}
  P.~P.~Kulish and E.~K.~Sklyanin,
  ``Quantum Spectral Transform Method. Recent Developments,''
  Lect.\ Notes Phys.\  {\bf 151} (1982) 61.
  %%CITATION = LNPHA,151,61;%%
  %71 citations counted in INSPIRE as of 21 May 20

%\cite{Tarasov:1986mc}
\bibitem{Tarasov}
  V.~O.~Tarasov,
  ``Irreducible Monodromy Matrices For The R Matrix Of The Xxz Model And
  %Local Lattice Quantum Hamiltonians,''
  Theor.\ Math.\ Phys.\  {\bf 63} (1985) 440
   [Teor.\ Mat.\ Fiz.\  {\bf 63} (1985) 175];
%\bibitem{Tarasov:1992aw}
  V.~Tarasov,
  ``Cyclic monodromy matrices for sl(n) trigonometric R matrices,''
  Commun.\ Math.\ Phys.\  {\bf 158} (1993) 459
  [hep-th/9211105].
  %%CITATION = HEP-TH/9211105;%%
  %10 citations counted in INSPIRE as of 21 May 2013


%\cite{Molev:1994rs}
\bibitem{Molev}
  A.~Molev, M.~Nazarov and G.~Olshansky,
  ``Yangians and classical Lie algebras,''
  Russ.\ Math.\ Surveys {\bf 51} (1996) 205
  [hep-th/9409025].
  %%CITATION = HEP-TH/9409025;%%
  %37 citations counted in INSPIRE as of 21 May 2013


%\cite{KRS}
\bibitem{KRS}
  P.~P.~Kulish, N.~Y.~.Reshetikhin and E.~K.~Sklyanin,
  ``Yang-Baxter Equation and Representation Theory. 1.,''
  Lett.\ Math.\ Phys.\  {\bf 5} (1981) 393.
  %%CITATION = LMPHD,5,393;%%
  %357 citations counted in INSPIRE as of 26 May 2013
%\cite{Faddeev:1996iy}
\bibitem{Faddeev96}
  L.~D.~Faddeev,
 ``How algebraic Bethe ansatz works for integrable model,''
  hep-th/9605187.
  %%CITATION = HEP-TH/9605187;%%
  %247 citations counted in INSPIRE as of 26 May 2013



\bibitem{Gelfand}
I.M. Gelfand and G.E. Shilov,
Generalized Functions. Vol. I: Properties
 and Operations, Boston, MA:  Academic Press, 1964.



\bibitem{BCFW}
  R.~Britto, F.~Cachazo and B.~Feng,
  ``New recursion relations for tree amplitudes of gluons,''
  Nucl.\ Phys.\ B {\bf 715} (2005) 499
  [hep-th/0412308]; \\
  %%CITATION = HEP-TH/0412308;%%
  %439 citations counted in INSPIRE as of 24 May 2013
  R.~Britto, F.~Cachazo, B.~Feng and E.~Witten,
  ``Direct proof of tree-level recursion relation in Yang-Mills theory,''
  Phys.\ Rev.\ Lett.\  {\bf 94} (2005) 181602
  [hep-th/0501052].
  %%CITATION = HEP-TH/0501052;%%
  %508 citations counted in INSPIRE as of 24 May 2013



\end{thebibliography}
\end{document}